\documentclass[sigplan,screen]{setting/acmart}
\usepackage{amsmath,amsfonts}
\usepackage{caption}
\usepackage{hyperref}
\usepackage{graphicx,tabularx}
\usepackage{textcomp}
\usepackage[frozencache,cachedir=.]{minted}
\usepackage{xcolor}
\definecolor{bgcolor}{rgb}{0.95, 0.95, 0.95} 
\newcommand{\artifact}{PPoPP25\_Artifact\_TurboFFT.zip}
\usepackage{xcolor}
\usepackage{booktabs}
\usepackage{tagging}
\usepackage{listings}
\usepackage{algpseudocode}

\usepackage[normalem]{ulem}

\usepackage[font=small,skip=0pt]{caption}
\usepackage{multicol}
\usepackage{multirow}

\usepackage{booktabs}   
\usepackage{subcaption} 

\usepackage{enumitem}
\usepackage{algorithm}
\usepackage{algpseudocode}
\usepackage{multicol}
\usepackage{siunitx} 
\usepackage{etoolbox}
\usepackage{soul}
\usepackage{cancel}
\usepackage[normalem]{ulem}

\usepackage[font=small,skip=0pt]{caption}

\usepackage{amsmath,tabularx}

\newcommand{\multiline}[1]{%
    \begin{tabularx}{\dimexpr\linewidth-\ALG@thistlm}[t]{@{}X@{}}
    #1
    \end{tabularx}
}

\setcopyright{acmcopyright}
\acmYear{2025}\copyrightyear{2025}
\setcopyright{licensedusgovmixed}
\acmConference[PPoPP'25]{ACM SIGPLAN Symposium on Principles and Practice of Parallel Programming}{March 01--05, 2025}{Las Vegas, NV, USA}
\acmYear{2025}
\begin{document}

\pdfcompresslevel=9
\pdfminorversion=5
\pdfobjcompresslevel=2


\title{TurboFFT: Co-Designed High-Performance and Fault-Tolerant Fast Fourier Transform on GPUs}

\author{Shixun Wu}
\affiliation{%
  \institution{UC Riverside}
  \city{Riverside}
  \state{CA}
  \country{USA}}
\email{swu264@ucr.edu}

\author{Yujia Zhai}
\affiliation{%
  \institution{UC Riverside}
  \city{Riverside}
  \state{CA}
  \country{USA}}
\email{yzhai015@ucr.edu}

\author{Jinyang Liu}
\affiliation{%
  \institution{University of Houston}
  \city{Houston}
  \state{TX}
  \country{USA}}
\email{jliu217@central.uh.edu}

\author{Jiajun Huang}
\affiliation{%
  \institution{UC Riverside}
  \city{Riverside}
  \state{CA}
  \country{USA}}
\email{jhuan380@ucr.edu}

\author{Zizhe Jian}
\affiliation{%
  \institution{UC Riverside}
  \city{Riverside}
  \state{CA}
  \country{USA}}
\email{zjian106@ucr.edu}

\author{Huangliang Dai}
\affiliation{%
  \institution{UC Riverside}
  \city{Riverside}
  \state{CA}
  \country{USA}}
\email{hdai022@ucr.edu}

\author{Sheng Di}
\affiliation{%
  \institution{Argonne National Laboratory}
  \city{Lemont}
  \state{IL}
  \country{USA}
  }
\email{sdi1@anl.gov}

\author{Franck Cappello}
\affiliation{%
  \institution{Argonne National Laboratory}
  \city{Lemont}
  \state{IL}
  \country{USA}
  }
\email{cappello@mcs.anl.gov}

\author{Zizhong Chen}
\affiliation{%
  \institution{UC Riverside}
  \city{Riverside}
  \state{CA}
  \country{USA}
  }
\email{chen@cs.ucr.edu}
\renewcommand{\shortauthors}{Wu et al.}
\begin{abstract}
  GPU-based fast Fourier transform (FFT) is extremely important for scientific computing and signal processing. However, we find the inefficiency of existing FFT libraries and the absence of fault tolerance against soft error. To address these issues, we introduce TurboFFT, a new FFT prototype co-designed for high performance and online fault tolerance. For FFT, we propose an architecture-aware, padding-free, and template-based prototype to maximize hardware resource utilization, achieving a competitive or superior performance compared to the state-of-the-art closed-source library, cuFFT. For fault tolerance, we 1) explore algorithm-based fault tolerance (ABFT) at the thread and threadblock levels to reduce additional memory footprint, 2) address the error propagation by introducing a two-side ABFT with location encoding, and 3) further modify the threadblock-level FFT from 1-transaction to multi-transaction in order to bring more parallelism for ABFT. Our two-side strategy enables online correction without additional global memory while our multi-transaction design averages the expensive threadblock-level reduction in ABFT with zero additional operations. Experimental results on an NVIDIA A100 server GPU and a Tesla Turing T4 GPU demonstrate that TurboFFT without fault tolerance is comparable to or up to 300\% faster than cuFFT and outperforms VkFFT. TurboFFT with fault tolerance maintains an overhead of 7\% to 15\%, even under tens of error injections per minute for both FP32 and FP64.
\end{abstract}


\maketitle
\section{Introduction}
The Fast Fourier Transform (FFT) is a core computation in a wide range of applications. For example, FFT is employed extensively in science and engineering, such as exascale projects like LAMMPS \cite{thompson2022lammps}, quantum simulations like QMCPACK \cite{kim2018qmcpack}, molecular dynamics like HACC \cite{habib2016hacc}. A significant portion, e.g. 70\%, of processing time in scientific applications is consumed by FFT, showcased by a space telescope project \cite{stockman1999ngst, murphy1998ngst}. However, those applications are increasingly vulnerable to transient faults caused by high circuit density,  low near-threshold voltage, and low near-threshold voltage \cite{lutz1993analyzing, nicolaidis1999time,laprie1985dependable}. Oliveira et al. \cite{oliveira2017experimental} demonstrated an exascale
system with 190,000 cutting-edge Xeon Phi processors
that still suffer from daily transient errors under ECC protection. Recognizing the importance, the U.S. Department of Energy has named reliability as a major challenge \cite{lucas2014doe}.

Since Intel Corporation first reported a transient error and the consequent soft data corruption in 1978  \cite{may1979alpha}, both the academic and industry sectors have recognized the significant impact of transient faults. In 2003, Virginia Tech broke down and sold online its newly-built Big Mac cluster of 1100 Apple Power Mac G5 computers because it lacked ECC protection, leading to unusability due to cosmic ray-induced crashes \cite{geist2016supercomputing}. Despite ECC protection, transient faults remain a threat to system reliability. For instance, Oliveira et al. simulated an exascale system with 190,000 cutting-edge Xeon Phi processors, finding it still vulnerable to daily transient errors under ECC \cite{oliveira2017experimental}. Such faults have not only been a concern in simulations; Google has encountered transient faults in its real-world production environment, resulting in incorrect outputs \cite{hochschild2021cores}. In response to the persistent challenge posed by transient faults on large-scale infrastructure services \cite{wu2024dgro,zukswarm}, Meta initiated an investigation in 2018 to find solutions \cite{dixit2021silent}. 

Transient faults can result in fail-stop errors, causing crashes, or fail-continue errors, leading to incorrect results. Checkpoint/restart mechanisms \cite{phillips2005scalable, NEURIPS2019_9015, tao2018improving, tensorflow2015-whitepaper} or algorithmic methods \cite{hakkarinen2014fail, chen2008scalable, chen2008extending, wu2024ft, wu2023anatomy, wu2023ft} can often mitigate fail-stop errors, whereas fail-continue errors pose a greater risk by silently corrupting application states and yielding incorrect result \cite{mitra2014resilience, cher2014understanding, dongarra2011international, calhoun2017towards, snir2014addressing}. These errors can be especially challenging in safety-critical scenarios \cite{li2017understanding}. In this paper, we concentrate on fail-continue errors occurring in computing logic units, such as incorrect outcomes, and assume that memory and fail-stop errors are addressed through error-correcting codes and checkpoint/restart. We describe them as soft errors.

To protect FFT against soft errors, a variety of fault tolerance methods have been proposed \cite{may1979alpha, baumann2002soft,geist2016supercomputing}. Jou and Abraham suggested an algorithm-based fault tolerance (ABFT) method for FFT networks that provides full fault coverage and throughput with a hardware overhead of $O(\frac{2}{\log_2N})$ \cite{jou1988fault}. Pilla et al. proposed a specific software-based hardening strategy to lower failure rates \cite{pilla2014software}. Fu and Yang implemented a fault-tolerant parallel FFT using MPI \cite{fu2009fault}. Xin et al. integrated a fault tolerance scheme into FFTW \cite{liang2017correcting}.

However, the fault tolerance for FFT on modern GPUs is absent. Besides, we find that offline FT-FFT suffers from significant overhead due to high kernel launch overhead and redundant memory operations. To hide the fault tolerance memory footprint, we fuse the ABFT operation at thread and threadblock levels. During our development at an FFT baseline for fused ABFT, we surprisingly characterized the inefficiency within the existing GPU-based FFT libraries, including the state-of-the-art closed-source cuFFT and the popular open-source VkFFT. Their sub-par performance mainly comes from underutilization due to poor kernel parameters and shared memory padding. After developing a high-performance FFT baseline, we further reduce the fused ABFT by assigning more than one global memory transaction for each threadblock. To address these issues, we propose TurboFFT \footnote{available at https://github.com/shixun404/TurboFFT.git},  a high-performance FFT implementation equipped with a two-sided fault tolerance scheme that detects and corrects silent data corruptions at computing units on-the-fly. More specifically, our contributions include:

\begin{itemize}[leftmargin=*]

\item \textbf{High-Performance FFT with Fault Tolerance:} We develop TurboFFT, a new FFT prototype co-designed for high performance and online fault tolerance. TurboFFT has: $1)$ performance comparable to or faster than state-of-the-art closed-source library, cuFFT, and $2)$ on-the-fly error correction with low overhead.

\item \textbf{Architecture-Aware, Padding-Free, Template-based:} Our design maximizes hardware utilization by 1) architecture aware optimizations such as memory coalescing, reorganizing global memory accessing in the last FFT stage, avoid FP64 triangular function with twiddling factor table, 2) replacing shared memory padding with swizzling access to avoid bank conflict, and 3) zero-cost kernel parameter tunning with template-based code generation for a wide range of problem sizes. 

\item \textbf{Two-side, Fused, Multi-Transaction ABFT:} We: 1) address the exponential error propagation by introducing a new two-side ABFT with location encoding, 2) explore ABFT with kernel fusion at the thread, warp, and thread block levels to avoid additional memory transaction, 3) further reduce the ABFT threadblock-level reduction overhead by changing FFT from 1-transaction to multiple transactions. Transaction means a round trip of \textit{global memory read} $\rightarrow$ \textit{threadblock-level FFT} $\rightarrow$ \textit{global memory write}.

\item  \textbf{Competitive for both FP32 and FP64, even under error injection:} Experimental results of single precision and double precision on an NVIDIA A100 server GPU and a Tesla Turing T4 GPU show that TurboFFT without fault tolerance offers a competitive or superior performance compared to the state-of-the-art closed-source library cuFFT. TurboFFT only incurs a minimum overhead (7\% to 15\% on average) compared to cuFFT, even under tens of error injections per minute for FP32 and FP64.

\end{itemize}

\begin{figure}[ht]
    \centering
    \includegraphics[width=1\linewidth]{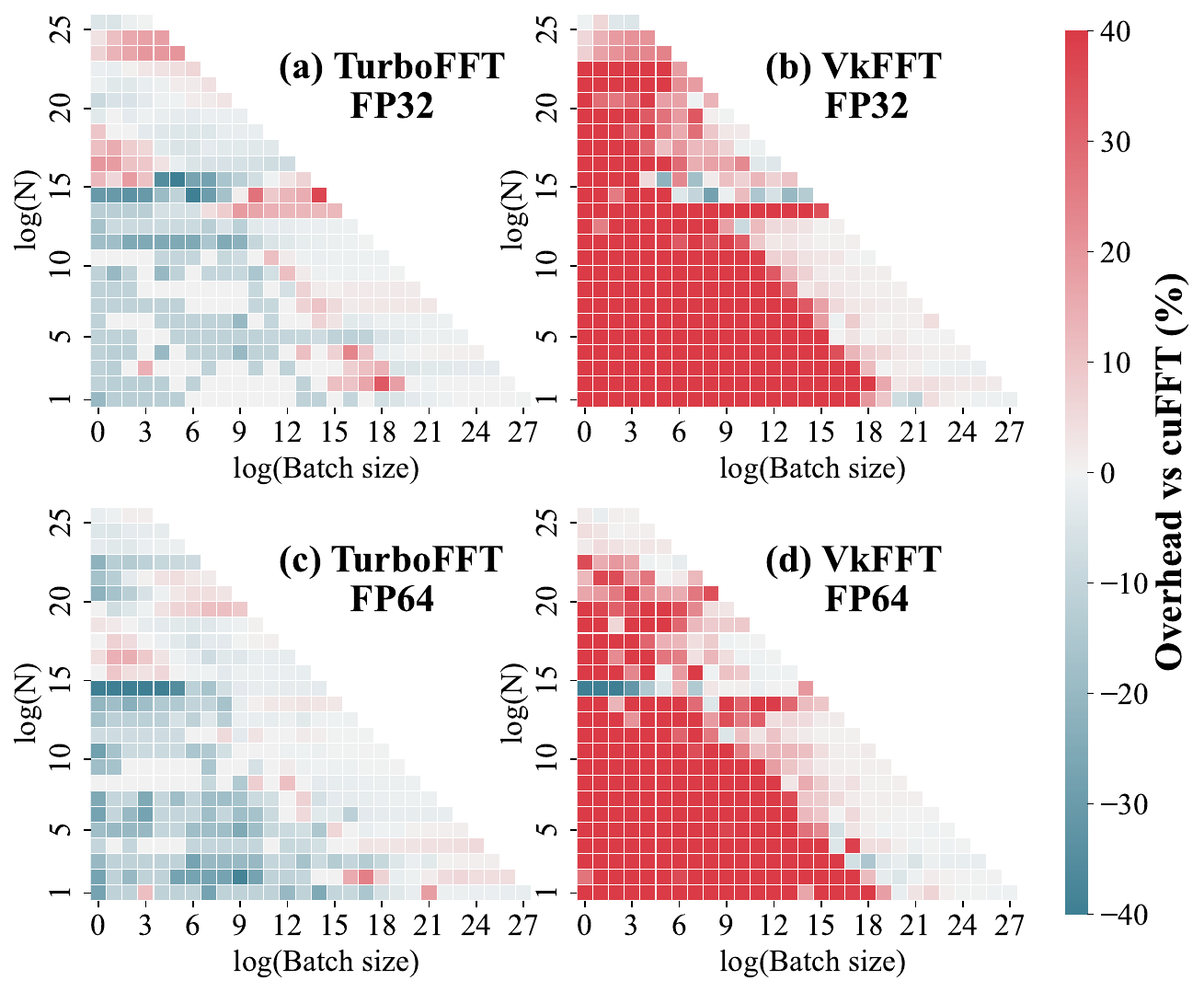}
    \caption{FFT performance comparison with cuFFT on an NVIDIA A100 GPU. Red means slower. Green means faster. Grey means close. TurboFFT is competitive with cuFFT and outperforms VkFFT.}
    \label{fig:TurboFFT_vs_VkFFT_vs_cuFFT}
\end{figure}
\section{Background}
\label{sec:background}


\subsection{Inefficiency of Existing FFT Libraries} According to our benchmark of TurboFFT and VkFFT against cuFFT in Figure \ref{fig:TurboFFT_vs_VkFFT_vs_cuFFT}, we find the following performance issues:
\subsubsection{Sub-par Performance of open-source library} VkFFT focuses on FP64, lacks optimizations on FP32, as well as small problem size, e.g. \textit{batch size} $\cdot N \cdot 8 b\leq 256 MB$, as shown in Figure \ref{fig:TurboFFT_vs_VkFFT_vs_cuFFT} (b) (d). In contrast, TurboFFT provides architecture-aware optimizations for both FP32 and FP64. Besides, we also present a padding-free store from register to shared memory by swizzling the store order between different threads.

\subsubsection{Underutilization of streaming processors (SMs)} The state-of-the-art closed library fails to fully utilize the SMs on A100 for problem sizes less than $2$ MB, demonstrated by the many green squares at the lower triangular area in Figure \ref{fig:TurboFFT_vs_VkFFT_vs_cuFFT} (a) and (c). TurboFFT outperforms cuFFT with better kernel parameters. The kernel parameters are manually searched with a cost-friendly template-based code generation strategy.

\subsection{Absence of Fault Tolerant FFT}
  
  \subsubsection{DFT and FFT} FFT accelerates discrete Fourier transform (DFT) computation through factorization, as shown in Figure \ref{fig:error_propagation}. The forward DFT maps a complex sequence $\mathbf{x}=(x_0, x_1,\cdots,x_{N-1})$ to $\mathbf{y}=(y_0,\cdots, y_{N-1})$, where 
$y_j$  $= \sum_{n = 0}^{N-1} x_n$ $e^{-2\pi ijn/N}$. The inverse discrete Fourier transform is defined as $x_j = \frac{1}{N}\sum_{n = 0}^{N-1} y_n e^{2\pi ijn/N}$. DFT can be treated as a matrix-vector multiplication (GEMV) between the DFT matrix $W=( \omega_N^{jk})_{j,k=0,\cdots,N-1}, \omega_N = e^{-2\pi i/N}$ and the input sequence $\mathbf{x}$. Given the expensive O($N^2$) operations of GEMV, FFT factorizes the DFT matrix into a sparse factors product and achieves a complexity of O($N\log N$) \cite{van1992computational}. 

\subsubsection{Exponential Error Propagation.} 
\begin{figure}[ht]
    \centering
    \includegraphics[width=1\linewidth]{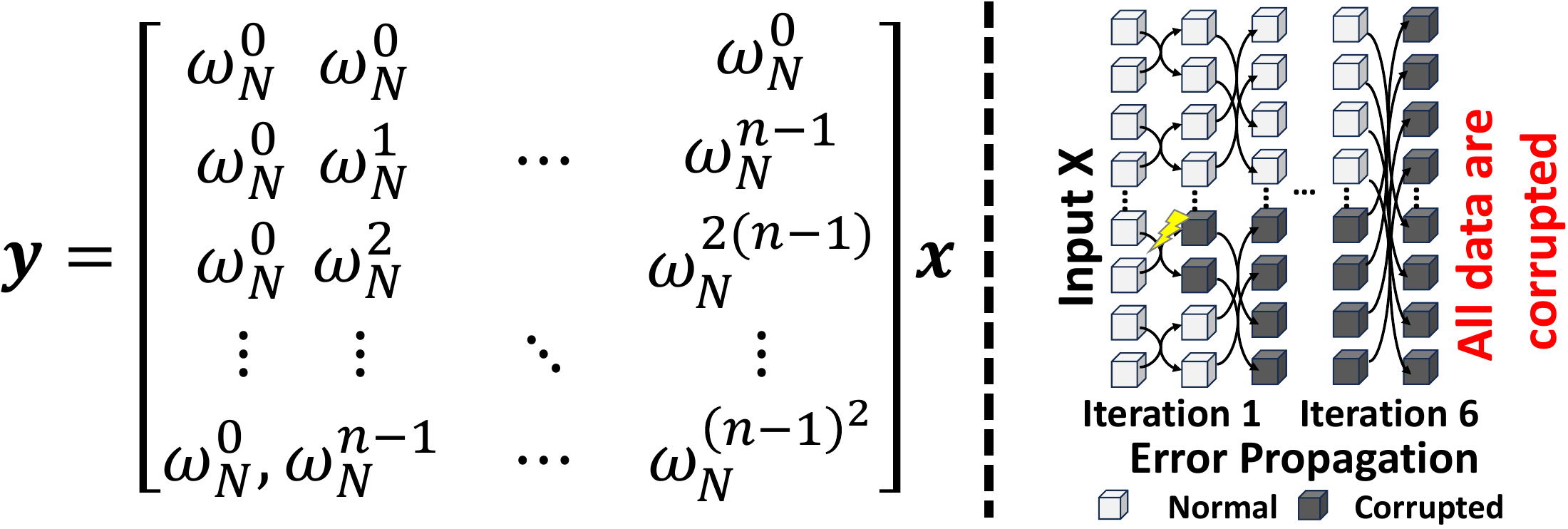}
    \caption{DFT and its Error Propagation}
    \label{fig:error_propagation}
\end{figure}

The butterfly operation doubles the number of corrupted data.  As shown in Figure \ref{fig:error_propagation}, one corrupted data in a 64-point FFT will spread to all other data in 6 iterations. Existing ABFT schemes detect errors by utilizing a perspective of GEMV in DFT, namely $(\mathbf{e}^TW)\mathbf{x}= \mathbf{e}^T(W\mathbf{x})$, as shown in Eqn. (\ref{eqn:ABFT}) and (\ref{eqn:C}).
\begin{equation}
    W\xrightarrow[]{encode}W^c:=\left[\begin{array}{c}W \\ \mathbf{e}^T W\end{array}\right],
    \label{eqn:ABFT}
\end{equation}

\begin{equation}
\mathbf{y}^c = W^c\cdot \mathbf{x} = \begin{bmatrix}W\mathbf{x} \\ \mathbf{e}^T W\mathbf{x} \end{bmatrix} = \begin{bmatrix}\mathbf{y} \\ \mathbf{e}^T\mathbf{y}
 \end{bmatrix},
\label{eqn:C}
\end{equation}
where $\mathbf{x}$ is the input signal, $X$ is the output signal, $W$ is the DFT matrix, and $\mathbf{e}$ is the encoding vector. The correctness is verified by comparing $(\mathbf{e}^TW)\mathbf{x}$ and $\mathbf{e}^T\mathbf{X}$. If $|(\mathbf{e}^TW)\mathbf{x} - \mathbf{e}^T\mathbf{X}|/|(\mathbf{e}^TW)\mathbf{x}|$ exceeds an error threshold $\delta$, it indicates that an error occurred during the computation. The selection of $\mathbf{e}$ is widely discussed because the 1's vector misses the opposite error $x+\epsilon$ and $x-\epsilon$ by addition. Jou \cite{jou1988fault} suggests $\mathbf{e}_{\text{Jou}} = (1/\omega_N^{-0},\cdots,1/\omega_N^{-(N-1)})$ while requiring a variant input $\mathbf{x}' = (2x_0+x_1,\cdots,2x_{N-1}+x_0)$, leading to additional computation overhead. Next, Wang \cite{wang1994algorithm} proposes $\mathbf{e}_{\text{Wang}} =  (\omega_3^0,\cdots,\omega_3^{N-1})$ allowing $\mathbf{x}$ to be unchanged. The computation of $\mathbf{e}^TW$ is not trivial and introduces addition computation or memory overhead. Due to the error propagation, existing fault tolerance schemes not only necessitate a checksum computation for each signal and a time-redundant recompute under error correction. In contrast,  TurboFFT employs a column vector $\mathbf{e}_c$ to linearly combine a batch of signals, as shown in Eqn. (\ref{eqn:TurboFFT_sec2}). The composite signal can recover the corrupted signal given the SEU assumption and the linearity of FFT, enabling batched detection and delayed correction without recomputation. The main contribution of our two-side ABFT is the performance improvement by amortizing one ABFT checksum transaction along a batch.
\begin{equation}
\textbf{Our 2-Side:~~}\mathbf{Y}^c = W^c\cdot \mathbf{Xe} = \begin{bmatrix}W\mathbf{Xe} \\ \mathbf{e}^T W\mathbf{xe} \end{bmatrix} = \begin{bmatrix}\mathbf{Ye} \\ \mathbf{e}^T\mathbf{Ye}
 \end{bmatrix},
\label{eqn:TurboFFT_sec2}
\end{equation}
\subsubsection{Significant Overhead of Offline ABFT} The work closest to ours is the offline FT-FFT presented in \cite{pilla2014software}. However, this method demonstrates high overhead. Hence, we further explore the kernel fusion of two-side ABFT on GPU to hide the memory footprint of checksum operations.  

\subsubsection{Fault Model} TurboFFT focuses on the detection and correction of errors at computing units that can affect the results of the final output. We assume memory errors are protected by ECC \cite{bird2017neutron} and the reliability issue of communication is protected by FT-MPI \cite{fagg2000ft}. To address the compute errors at the run-time,  we adopt a fault-tolerant scheme under a single-event upset (SEU) assumption \cite{binder1975satellite,petersen2013single,binder1975satellite}, i.e., there is only one soft error in each error detection and correction period. The SEU assumption is validated by the low occurrence rate of multiple soft errors caused by short fault detection intervals, widely used in many works \cite{reis2005swift, zhai2021ft, ding2011matrix, wu2014ft}. 

\section{TurboFFT without Fault Tolerance}
\label{sec:design_and_optimizations}
A high-performance fault-tolerant FFT implementation necessitates fusing the memory costs of fault detection with the original FFT operations. Hence, an efficient FFT framework serves as the foundation for memory operations of fault detection. As cuFFT is closed-source, implementing a high-performance FFT from scratch becomes inevitable. In this section, we present the step-wise optimizations of FFT. The optimizations include avoiding bank conflict, maximizing the L1 cache hit rate, reducing triangular operations, and a template-based code generation for parameter selection.
\subsection{Architecture-Aware Optimizations}
\begin{figure}[ht]
    \vspace{-3mm}
    \centering
    \includegraphics[width=0.9\linewidth]{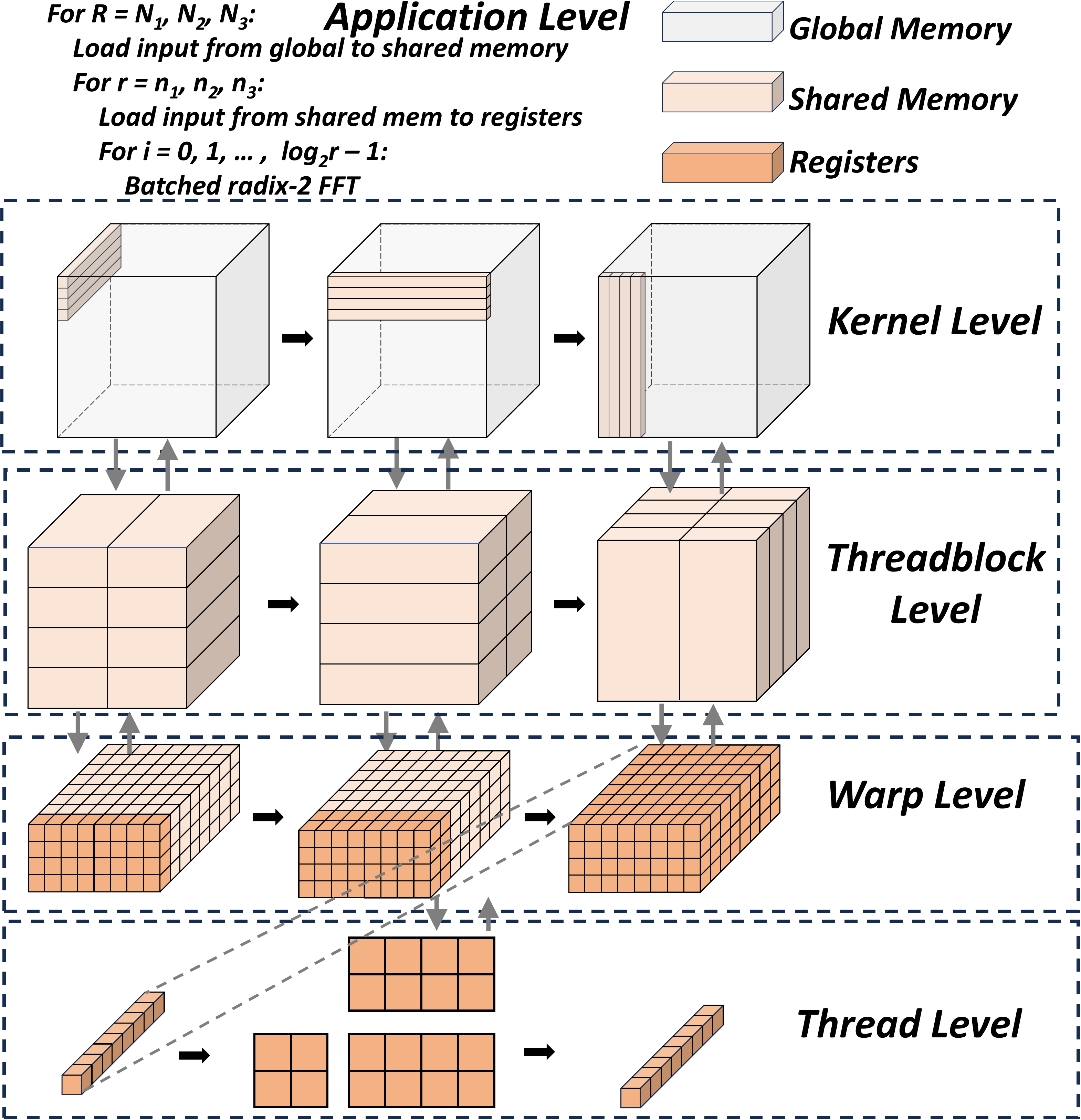}
    \caption{Overview of the optimized TurboFFT.}
    \vspace{-1mm}
    \label{fig:turbo_fft_details}
\end{figure}
TurboFFT architecture-aware optimizations including common tiled FFT, twiddling factor optimization, and global memory access optimizations. Figure \ref{fig:turbo_fft_details} demonstrates an overview of our optimized TurboFFT pipeline for FFT with large inputs, utilizing the step-wise optimizations mentioned above. Starting from the application level, the input signal is first tiled into a $N_1\times N_2\times N_3$ cube to fit into the shared memory maximum size of a threadblock. Next in the kernel level,  three stages of FFTs are performed step by step along each axis. For each stage, the FFT workload will be executed by threadblock, and each threadblock is assigned to a batch of FFT along the axis. Then at the threadblock level, the batched FFT loaded from global memory will first be arranged into a cube as the kernel level did. The cube size is equal to the thread-level radix. After that, warp-level FFT will load the input from shared memory into thread-level registers layer by layer. The shared memory access is coalesced to avoid bank conflicts and maximize the shared memory bandwidth. Once the thread-level inputs have been loaded into registers, the thread-level FFT macro kernel will perform the FFT computations. Within each level, the workflow goes from left to right. If a particular level's workload is completed, the tasks are then pop back to the higher level.
\begin{figure}[ht]
    \centering
    \includegraphics[width=0.9\linewidth]{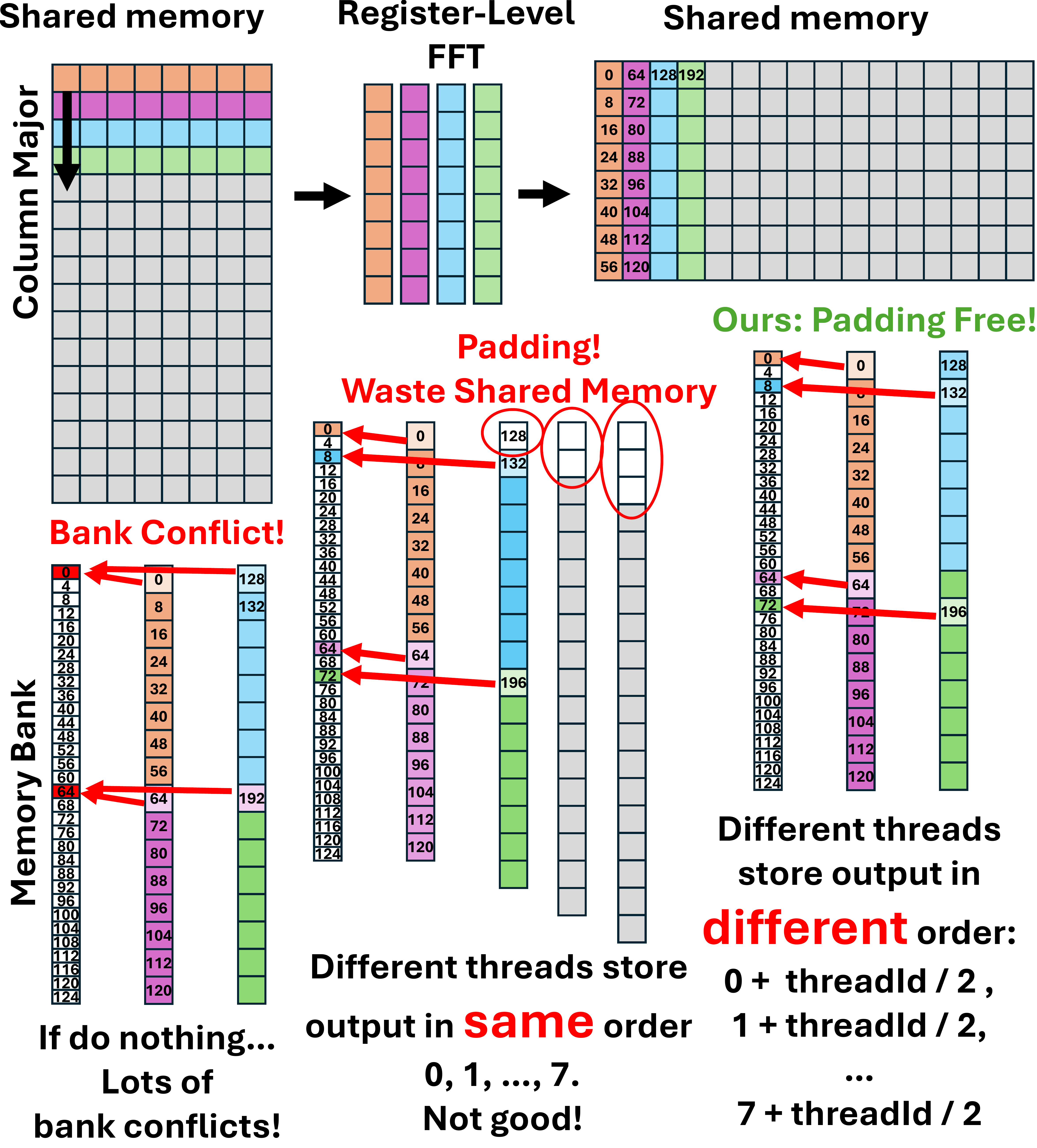}
    \caption{Our padding-free design achieves 100\% shared memory utilization with zero bank conflicts by rendering register store.}
    \label{fig:padding_free}
    \vspace{-1mm}
\end{figure}
\subsection{Padding-Free Store: Register to Shared Memory}
Figure \ref{fig:padding_free} demonstrates how our padding-free design swizzles the order of register to shared memory between different threads to avoid conflict without wasting shared memory. GPU-based FFT typically adopts the Stockham variant. In the Stockham variant, continuous thread access contiguous memory. For example, Figure \ref{fig:padding_free} shows a 128-point FFT decomposed into size $16\times 8$. First, a batch of $8$-point FFT is performed. Assuming the column-major store and data type is FP32, continuous elements in the $8$-point signals (row with the same color) have a stride of $16$. The loading from shared memory to register has zero bank conflict because the continuous access among threads naturally occupies the 128-byte bandwidth of the shared memory bank. After each thread finishes the FFT computation, the output signals should be stored back to shared memory for synchronization and adaptation to next step. However, each output signal will occupy a contiguous memory array, resulting in bank conflicts. To avoid bank conflict, VkFFT adds an 8-byte padding of every 128-byte data. Despite this method only introducing a low overhead of $8 / 128 = 6.25\%$, the irregular shared memory size with padding results in up to 40\% overhead due to the undividable shared memory capacity of a streaming processor, as shown in Figure \ref{fig:3xbarchart_overhead_A100_FP32}
and Figure \ref{fig:3xbarchart_overhead_A100_FP64} support our statement.
\subsection{Template-based Codegen}
A hard-coded FFT kernel can degrade performance when applied to different input sizes. However, rebuilding an FFT kernel with different parameter settings from scratch each time is not practical due to the high development cost. Generally, each FFT kernel in TurboFFT has around 2000 to 3000 lines of code (LOC) while an FFT with a large input size requires 3 different FFT kernels as mentioned above. To mitigate the increment of development cost, we propose a template-based code generation scheme to generate a series of input-size specific kernels. The code generation strategy we've developed involves utilizing semi-empirically chosen kernel parameters, tailored to input shapes, to generate highly efficient, parameterized kernels in real time. 
\begin{table}[ht] \centering
\caption{TurboFFT kernel parameter setup on Tesla T4.}
\begin{tabular}{l
    S[table-format=3] 
    S[table-format=3] 
    S[table-format=3] 
    S[table-format=3] 
    S[table-format=3] 
    S[table-format=3] 
    S[table-format=3]
    }
\toprule
            {$N$}        & {$N_{1}$}       & {$N_{2}$}       & {$N_{3}$}  & {$n_1$} & {$n_2$} & {$n_3$} & {$bs$} \\ 
\midrule
$2^{10}$              & {$2^{10}$}          & {$$}              & {$$}    & {$8$} & {$$}    & {$$}   &  {$1$} \\ 
$2^{17}$              & {$2^8$}         & {$2^9$}             & {$$}      & {$16$} & {$16$}    & {$$} & {$8$}\\ 
$2^{23}$    & {$2^8$}         & {$2^7$}             & {$2^8$}      & {$16$}  & {$16$}   & {$16$} &{$16$} \\  
\bottomrule
\end{tabular}
\label{tab:kernel_size}
\end{table}

\subsubsection{Code Generation Strategy} Besides following the step-wise TurboFFT optimization, the code generation scheme takes 7 parameters as input and generates a corresponding high-performance FFT kernel. The kernel parameters are $N_1, N_2, N_3$, $n_1, n_2, n_3$ and $bs$. $N_1, N_2, N_3$ are the cube size of the kernel-level input signal. $n_1, n_2, n_3$ are the cube size of the threadblock-level input signal. $bs$ denotes the number of FFT signals a thread will take for one computation. In the code generation template, the memory operations are strategically designed to sidestep bank conflicts. Additional parameters, such as the data type for vectorized load/store operations, are determined directly by threadblock-level input sizes and batch size. Figure \ref{alg:codegen} shows our code generation template.

\subsubsection{Kernel Parameters}
We generated FFT kernels with various parameters using the code generation template. Through empirical analysis, we identified a series of best kernels for input shapes from $2^3$ to $2^{29}$ and batch size from $1$ to $1024$, respectively. For input size in  $0\sim2^{13}$, $2^{14}\sim2^{22}$, and $2^{23}\sim2^{29}$, we adopt one, two, and three FFT kernel launches, respectively. Table \ref{tab:kernel_size} shows kernel parameters of 3 kernels.

\section{TurboFFT with Fault Tolerance}
\subsection{Two-sided Checksum}
Figure \ref{tab:illustrtation} details our motivation and algorithm designs. Previous FT-FFT, either offline or online, requires \textit{per-signal checksum}, namely applying the encoding vector to each input. Other than that, previous work states that a time-redundant recomputation is necessary. Motivated by the error propagation issue, we find the single error in FFT, although it corrupts hundreds of elements, the errors in the output are proportional to the initial error, as shown in the blue region in Figure \ref{tab:illustrtation}. To avoid frequent checksum encoding, we employ another encoding vector $\mathbf{e}_3=(1,2,\cdots,N)$ to \textit{linear combine} a batch of inputs within each thread and encode those thread-local variables with $\mathbf{e}_1$ at a fixed interval. The corrupted input can be located using the location encoding as shown in the green region of Figure \ref{tab:illustrtation}.

 \begin{figure}[ht]
     \centering
     \includegraphics[width=\linewidth]{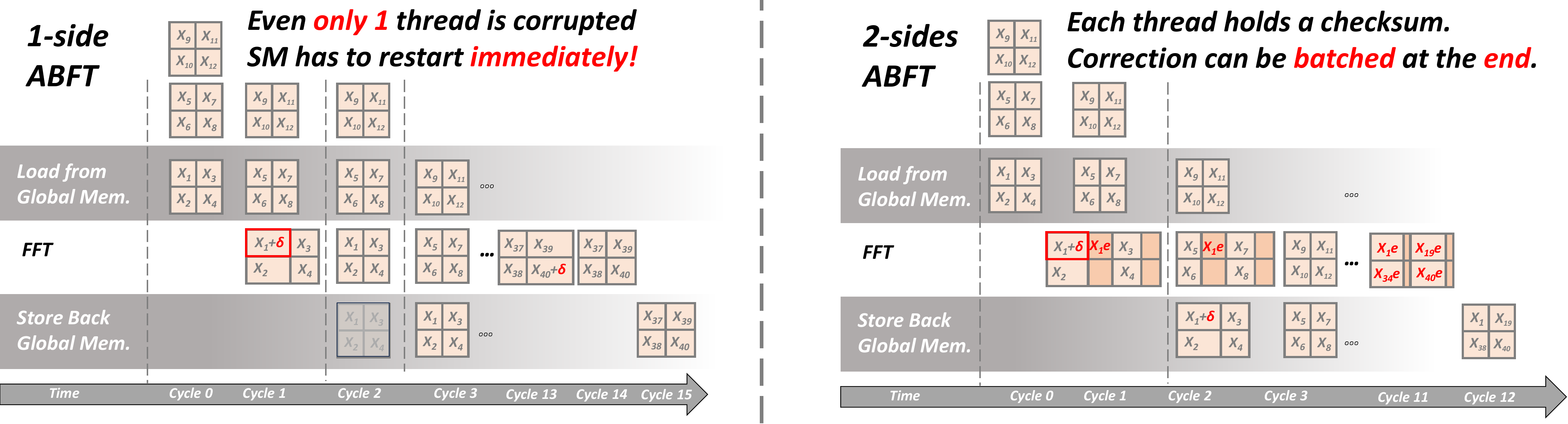}
     \caption{Benefit of two-sided ABFT: Delayed Batched Correction }
     \label{fig:why-two-sided-abft}
 \end{figure}



\subsubsection{Error Dectection}

 \begin{equation}
    X\xrightarrow[]{two-sided}X^c:=\begin{bmatrix}X &  X\mathbf{e}\end{bmatrix},
    \label{eqn:two_side_X}
\end{equation}
\begin{equation}
Y^c = W^c\cdot X^c = \begin{bmatrix}WX & WX\mathbf{e} \\ \mathbf{e}^T W\mathbf{x} & \mathbf{e}^TWX\mathbf{e} \end{bmatrix} = \begin{bmatrix}Y & Y\mathbf{e} \\ \mathbf{e}^TY & \mathbf{e}^TY\mathbf{e}
 \end{bmatrix},
\label{eqn:two_side_Y}
\end{equation}
There are two checksums being computed in the two-sided ABFT, namely the left-side $eWX$ and the right-side $WXe$. The checksum along the individual signal, namely one column in $X$, is used to detect the location of the error. Once a mismatch between the checksum before and after one FFT process, it is treated as an error. If an error is detected, then the divergence of the checksum along different signals between before and after the FFT is added back to the detected signal to perform the correction. According to Figure \ref{tab:illustrtation}, both two-sided ABFT and one-sided ABFT compute the checksum on the left side, i.e., eWX. In one-sided ABFT, this checksum is used to detect whether an error has occurred, and if an error is detected, the process reverts to a previously saved state for recalculation. In two-sided ABFT, while using this left-side checksum, the sum is also taken along the row direction of the current data X. If we perform an FFT operation on this checksum vector and subtract it from the checksum of the output result, we can then obtain the correction value for the entire column of erroneous data. Next, we just need to add this correction value back to correct the error.

 \subsubsection{Delayed Batched Correction}
\begin{figure}[ht]
    \centering
    \includegraphics[width=\linewidth]{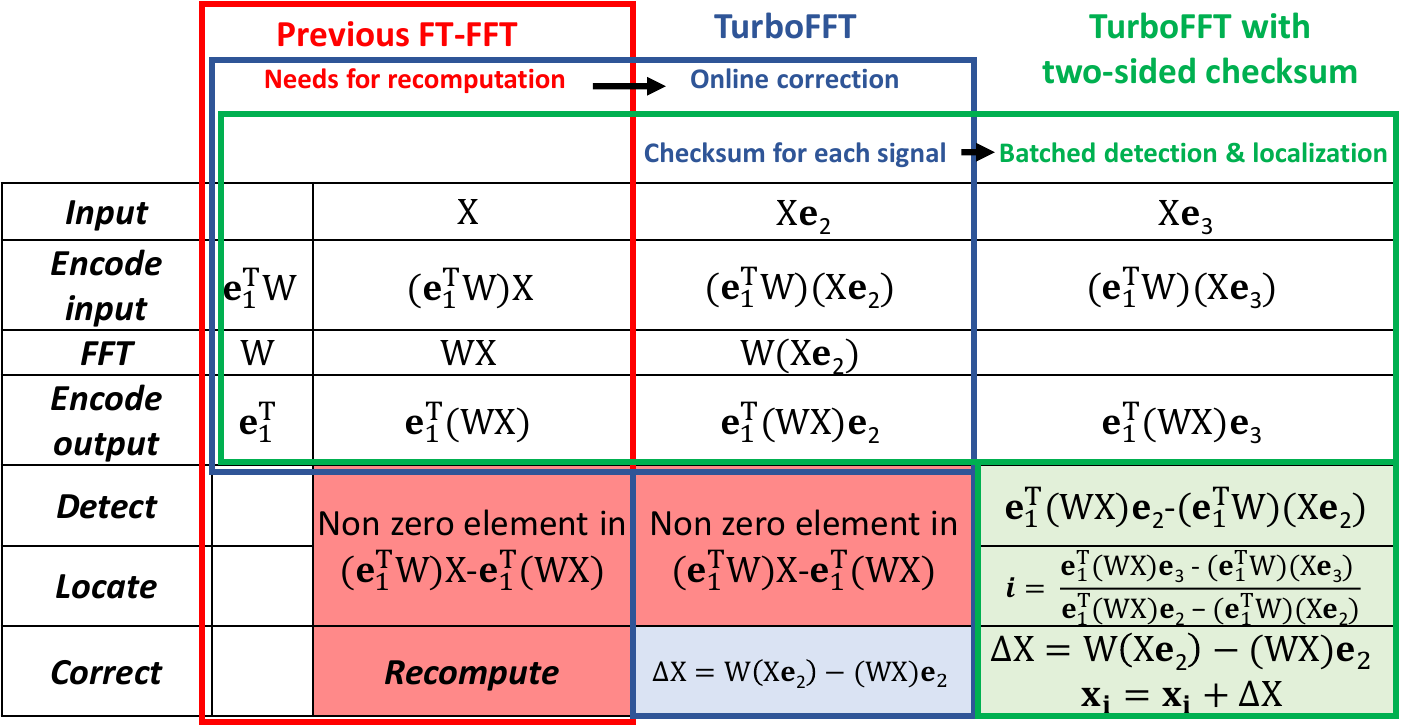}
    \caption{Motivation: Red area incurs high overhead. Blue region enables online correction. Green region enables batched detection.}
    \label{tab:illustrtation}
\end{figure}

The biggest difference between one-sided ABFT and two-sided ABFT is whether there is an immediate need for recalculation. From Fig. \ref{tab:illustrtation}, we understand that two-sided ABFT requires the calculation of an additional set of checksums, and correction also involves performing an FFT operation on the checksum vector. So, compared to one-sided ABFT, where does the advantage of two-sided ABFT lie? This work points to the advantage, namely delayed batched correction.

Under the assumptions of ABFT, a single checksum can correct one error. Therefore, for two-sided ABFT, it is only necessary to note the position of the error, i, and then continue processing the data for position i+1, until the operation ends or a new error occurs. In those cases, we need to correct the contents of the checksum at the erroneous position $i$. As a result, there is an opportunity for batch-processing operations among different threads, which enhances parallelism. Moreover, since there is no need to stop and execute immediately, the running pipeline of the program is not affected, thereby avoiding stalls.
 Figure \ref{fig:why-two-sided-abft} illustrates the difference between one-sided ABFT and two-sided ABFT. Any fault tolerance mechanism is essentially a tradeoff between resources. Compared to one-sided ABFT, two-sided ABFT actually uses additional computation to reduce memory overhead. If one-sided ABFT chooses not to recalculate immediately, it will need to reload data from the storage device during the next computation. In contrast, two-sided ABFT \textit{compresses} the data already read into local registers in the form of checksums. Therefore, before a new error occurs, the error information can be decompressed at any time by the checksum and used to correct previous errors. 

Figure \ref{fig:why-two-sided-abft} demonstrates the capability of batch delay correction. The 2$\times$2 grid in the figure represents a warp. When thread 1 in a warp encounters an error, the entire warp can continue to operate normally until the program terminates or a new error occurs. In contrast, one-sided ABFT requires immediate re-execution, otherwise it would need to reload data from memory. When an error occurs, a thread can note that an error has occurred at this point. Error correction operations are only carried out when the next error occurs or when the program terminates. Placing the right-side checksum at the end of the loop is to prevent a potential second error from contaminating the checksum that has already recorded one error.

\subsection{Fused ABFT at Thread and Threadblock levels}
\label{sec:fault_tolerant}

The lightweight, high-performance TurboFFT provides us with an efficient framework to build lightweight fault-tolerant schemes from silent data corruption. This section introduces how we optimize the two-sided ABFT schemes in 4 steps. In summary, we first implemented an offline version based on cufft and cuBLAS. We then realized that SGEMV requires a single thread to traverse all batches or every element of a single signal. As shown in Figure \ref{tab:illustrtation}, offline ABFT, due to the need for the checksum of all the data, actually doubles the memory transactions. Therefore, the overhead for both one-sided and two-sided ABFT is very close to 100\%. To avoid such a huge memory overhead, we decided to reduce the workload of each thread. This meant dividing the original signal into certain batch sizes so that each thread's task shifted from computing the checksums of all batches to just a small part of them. Next, we connected our customized kernel with TurboFFT. However, the additional checksum still brought significant extra memory access. Ultimately, we found that we could fully fuse the ABFT checksum within a single thread, thereby reducing the additional memory operations to zero. Since the initial steps merely involve calling library functions and are quite straightforward, we proceed directly to the discussion of kernel fusion.
\begin{figure}[ht]
    \centering
\includegraphics[width=0.9\linewidth]{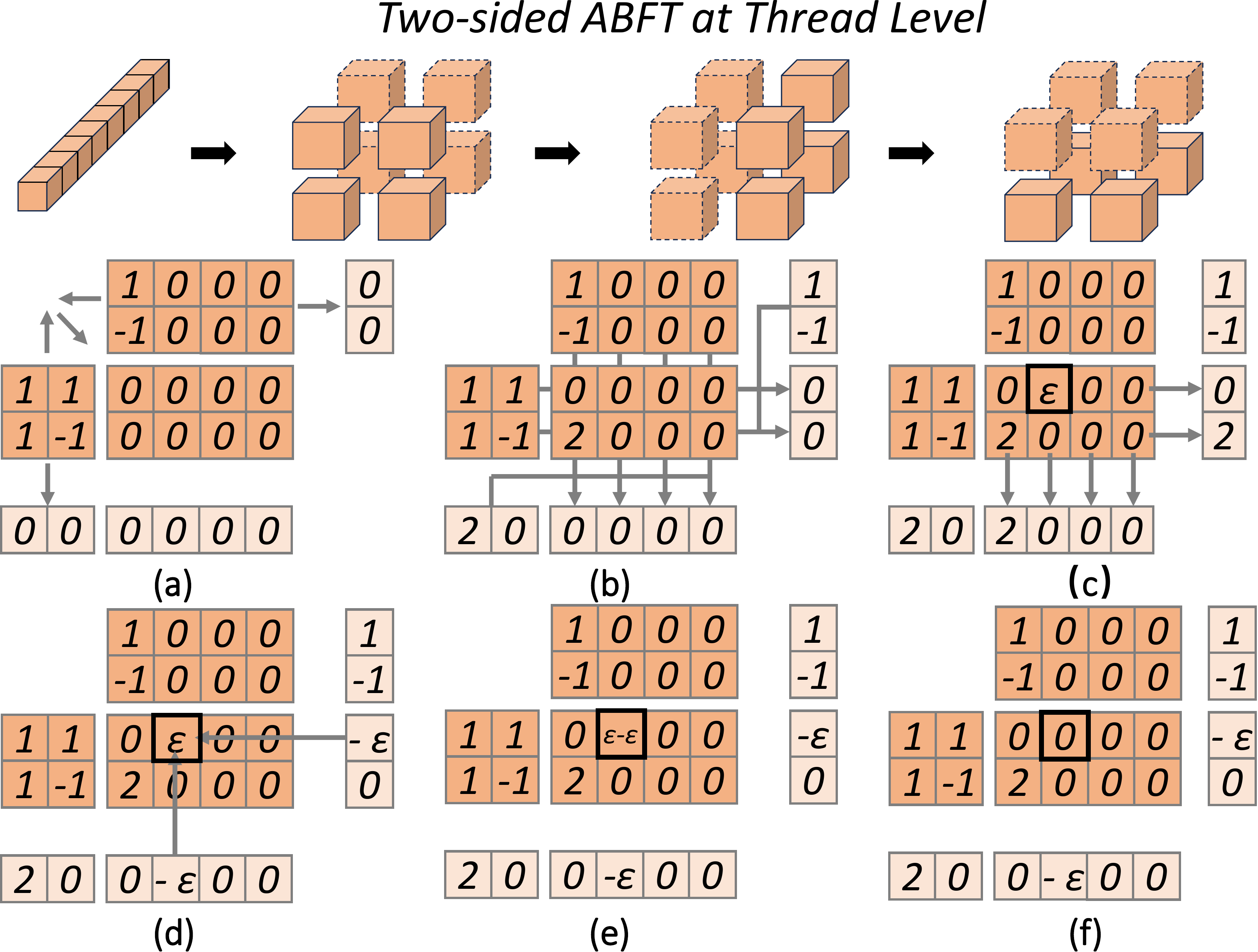}
    \caption{Thread-level Two-sided ABFT.}
    \label{fig:cft}
    \vspace{-1mm}
\end{figure}
\subsubsection{Thread-level Two-sided ABFT}
In Figure \ref{fig:cft}, we present how to use the two-sided ABFT to protect TurboFFT at the thread level. As shown in Figure \ref{fig:cft} (a), right-side checksums of the radix-2 DFT matrix and input signals are encoded while the batched FFT is being computed. Next in Figure \ref{fig:cft} (b), the row and column checksums of output signals are computed through the matrix-vector multiplication between DFT matrix/input signals and their encoded checksums. If there is an error $\epsilon$ in Figure \ref{fig:cft}(c), then the row and column checksums of output signals from reduction will hold a disagreement of $-\epsilon$ with the previous checksums. After that, the error can be further located in Figure \ref{fig:cft}(d)  and corrected with the disagreement value in Figure \ref{fig:cft}(e). Utilizing the ABFT scheme, the in-register computation will be protected from silent error. Figure \ref{alg:codegen} demonstrates the pseudocode of the computation fault-tolerant scheme. Although it does not introduce additional memory overhead, the redundant computation leads to significant overhead.
\begin{figure}[ht]
    \centering
\includegraphics[width=0.9\linewidth]{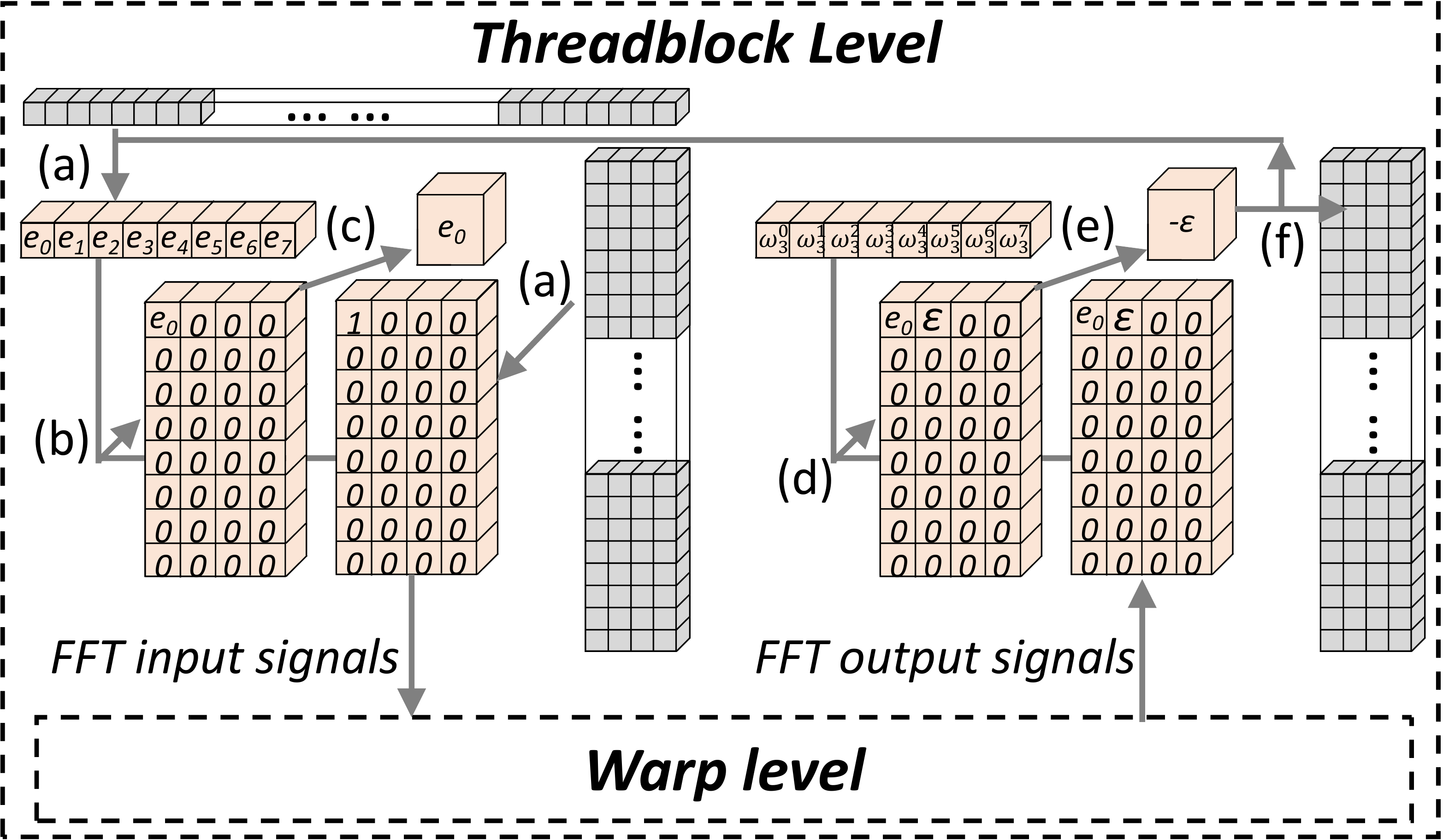}
    \caption{Checksum encoding at Threadblock Level.}
    \label{fig:s}
    \vspace{-1mm}
\end{figure}

\subsubsection{Threadblock-level Two-sided ABFT} In the following, we detail the threadblock-level two-sided ABFT. We protect the FFT at the threadblock level, as shown in Figure \ref{fig:s}. In TurboFFT, the input of threadblock-level FFT is batched of signals, and then each thread first performs the right-side ABFT, namely a vector addition, where the vector is the input signals from the global memory. No additional memory transaction is required. The encoding vector of the left-side checksum $e^TW$ is precomputed outside the FFT application and loaded into the threadblock from global memory into the shared memory, as shown in \ref{fig:s} (a). The input signal encoding is performed through register reuse, which is in conjunction with loading input signals from global memory. After that, the DFT checksum is obtained using CUDA warp shuffle primitives. Each thread then keeps a checksum of the DFT result in Figure \ref{fig:s} (b), (c). Next, the register stores the original inputs into the shared memory to wait for the execution of warp-level FFT. When the output signals are returned from warp level, one ABFT encoding is performed again to verify the correctness of the output signals. The access to the encoding vector does not occupy additional memory bandwidth while only requiring a minimal overhead of warp shuffling, as shown in Figure \ref{fig:s} (d), (e). If an error is detected in Figure \ref{fig:s} (f), the corresponding thread will record the location (batch ID of the input signal). Then the output signals will be uploaded back to global memory for subsequent workloads.

\subsection{Multi-Transaction Threadblock-level ABFT}
We further increase the batch size of threadblock-level ABFT by extending the workload of each threadblock from one transaction to multiple transactions. One transaction is a round-trip starting with reading input from global memory, then performing a threadblock-level FFT, and finally storing the data back to the global memory. In original FFT plans, a threadblock finishes once its transaction finishes. However, we extend the number of transactions from 1 to 2, 4, 8, 16, and 32. By switching to a multi-transaction plan, the workload of ABFT remains the same, namely a threadblock-level reduction at the end of transactions. A minimal thread-level accumulation is introduced to aggregate the input signals. The location encoding still applies, e.g. each thread aggregates the product of its share of the threadblock-level signal and the global ID for the signal. With the encoding, we can decode which signal is corrupted during error detection and then add the corrected value back to get the correct result.

\begin{figure}[ht]
    \centering
    \includegraphics[scale=0.22]{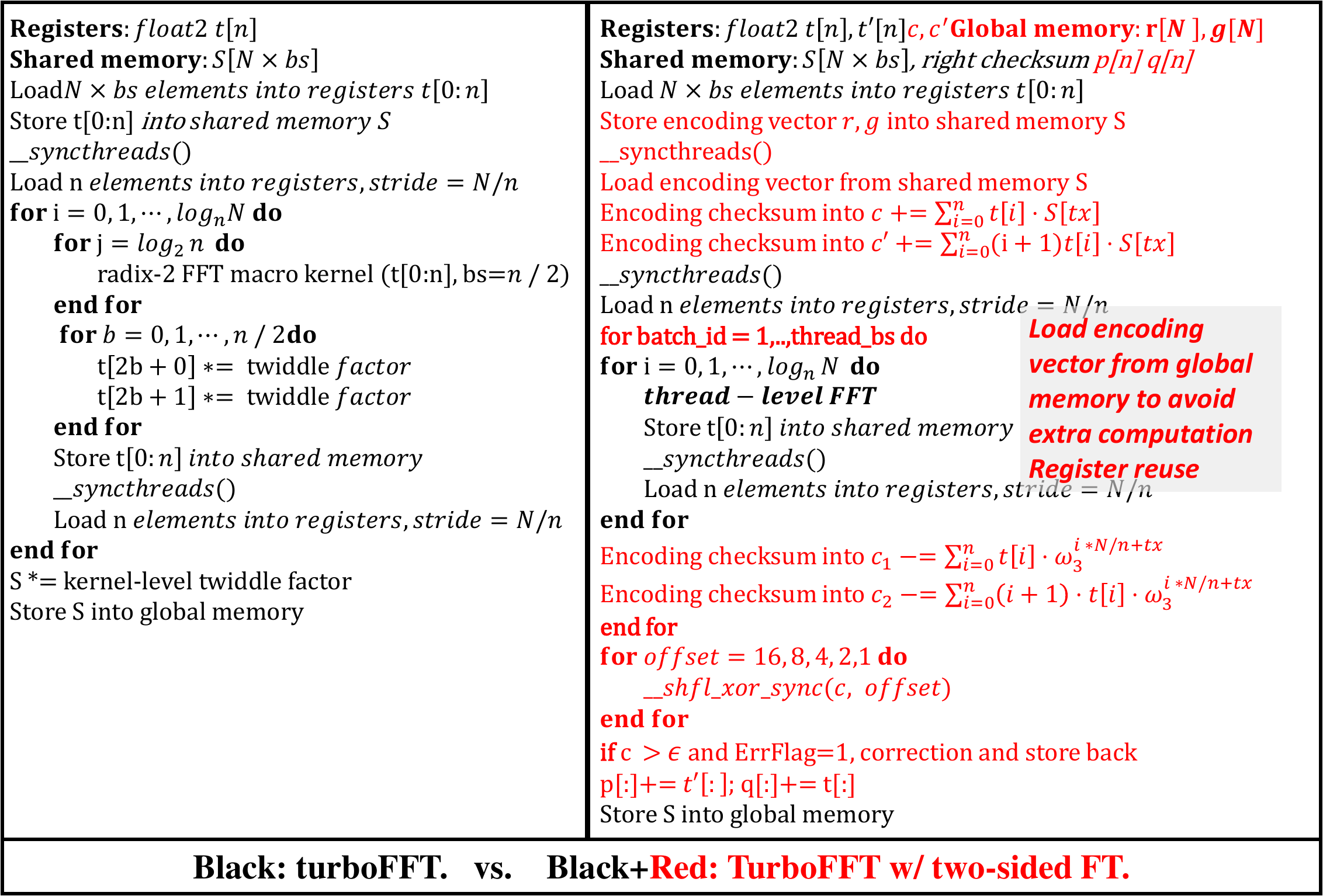}
      \caption{Codegen pseudocode: TurboFFT vs TurboFFT w/ FT.}
    \label{alg:codegen}
    \vspace{0mm}
\end{figure}

\section{Performance Evaluation}
\label{sec:results}
We evaluate TurboFFT on two NVIDIA GPUs, a Tesla Turing T4 and a 40GB A100-PCIE GPU. The Tesla T4 GPU is connected to a node with two 16-core Intel Xeon Silver 4216 CPUs, whose boost frequency is up to 3.2 GHz. The associated CPU main memory system has a capacity of 512 GB at 2400 MHz. The A100 GPU is connected to a node with one 64-core AMD EPYC 7763 CPU with a boost frequency of 3.5 GHz. We compile programs using CUDA $\mathtt{11.6}$ with the $\mathtt{-O3}$ optimization flag on the Tesla T4 machine, and using CUDA $\mathtt{12.0}$ on the A100 machine.
\subsection{TurboFFT without Fault Tolerance}
\begin{figure}[ht]
    \centering
    \includegraphics[width=1\linewidth]{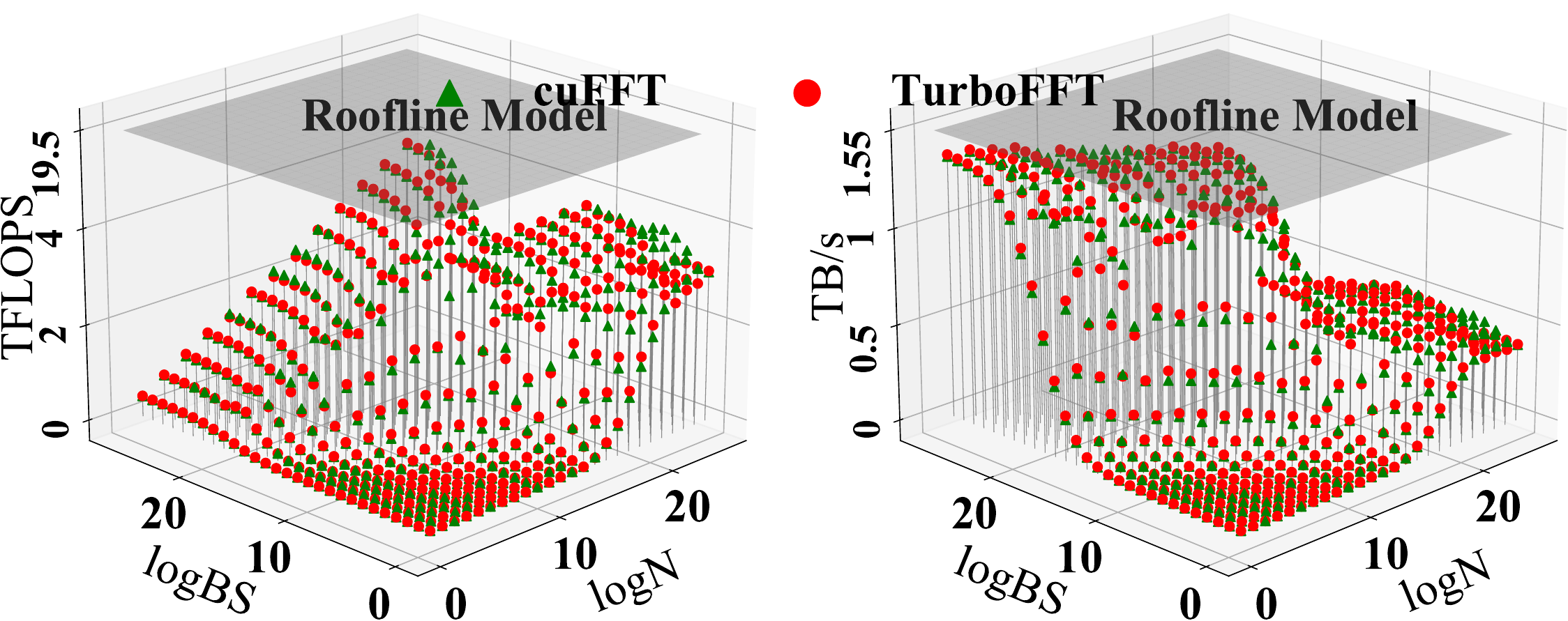}
    \caption{Performance of generated FP32 FFT kernels on A100}
    \label{fig:bench_FP32_a100}
    \vspace{-2mm}
\end{figure}

TurboFFT is faster than VkFFT and performs comparable or better than the state-of-the-art closed-source library cuFFT, as shown in Figure \ref{fig:TurboFFT_vs_VkFFT_vs_cuFFT}, \ref{fig:bench_FP32_a100}, \ref{fig:bench_FP64_a100},\ref{fig:3xbarchart_overhead_A100_FP32}, \ref{fig:3xbarchart_overhead_A100_FP64}, and \ref{fig:a100_bench}. 
We first analyze TurboFFT performance from the overhead overview in Figure \ref{fig:TurboFFT_vs_VkFFT_vs_cuFFT}. Then we detail our efforts on architecture-aware optimizations, padding-free design, and parameter tuning with Figure \ref{fig:3xbarchart_overhead_A100_FP32} and Figure \ref{fig:3xbarchart_overhead_A100_FP64}. Figure \ref{fig:3xbarchart_overhead_A100_FP32} and Figure \ref{fig:3xbarchart_overhead_A100_FP64} present a close observation of TurboFFT, VkFFT, and cuFFT from the perspective of fixed batch size and fixed signal length $N$. Figure \ref{fig:a100_bench} benchmark TurboFFT with VkFFT and cuFFT from the perspective of fixed problem size, $2$ GB for FP32 and $4$ GB for FP64.

\subsubsection{Overview of TurboFFT performance.}
Figure \ref{fig:TurboFFT_vs_VkFFT_vs_cuFFT} demonstrates the overhead compared to cuFFT with a heatmap. The heatmap includes overhead against cuFFT of problem size within $2$ GB for FP32 and $4$ GB for FP64. Each square represents a data point ($\log_2 N$, $\log(\text{Batch Size})$), where $N$ is the signal length. Red means the method is slower than cuFFT. Green means the method is faster than cuFFT. Grey means the method is close to cuFFT. As shown in Figure \ref{fig:TurboFFT_vs_VkFFT_vs_cuFFT}, the grey and green region of TurboFFT exceeds 90\%. For both FP32 and FP64, TurboFFT is faster than cuFFT for problem sizes within 2 MB. TurboFFT outperforms cuFFT by 40\% to 200\% on signal length $N = 2^{14}$. For FP64, TurboFFT outperforms cuFFT for signal length $2^{20}$ to $2^{21}$ as well. In contrast, VkFFT has more than 60\% red area (slower) compared to cuFFT.
\subsubsection{Architecture-Aware Optimizations.} 
\begin{figure}[ht]
    \centering
    \includegraphics[width=1\linewidth]{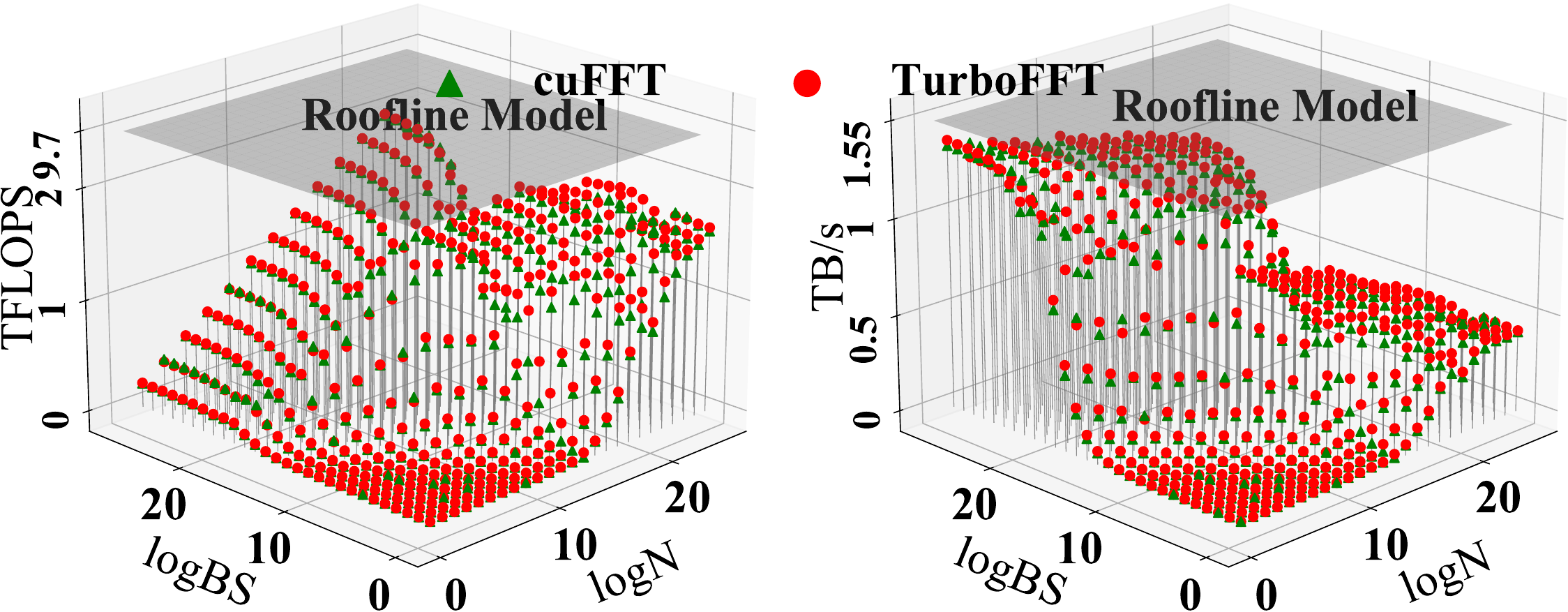}
    \caption{Performance of generated FP64 FFT kernels on A100}
    \label{fig:bench_FP64_a100}
\end{figure}

As shown in 
 Figure \ref{fig:bench_FP32_a100} and Figure \ref{fig:bench_FP64_a100}, TurboFFT achieves 90\% peak memory bandwidth. The memory bandwidth is measured by $2\times$ problem size divided by execution time. The bandwidth degradation at signal lengths $N 
 \geq 2^{14}$ is due to multiple kernel launches, typically two kernel launches for $2^{14} \leq N\leq2^{22}$, and three kernel launches for $N>2^{22}$. Hence the memory bandwidth is divided by the number of kernel launches. Our architecture-aware optimizations address the performance bottlenecks in computation and global memory transactions. Below is a detailed analysis of computation and global memory transaction bottleneck.

\textit{Computation} The computation bottleneck is primarily due to slow clock cycles caused by trigonometric functions or double-precision operations. Therefore, we pre-calculate and store the required trigonometric function values in global memory. The workload for each thread can be configured in our code generation strategy.
\begin{figure}[ht]
    \vspace{-2mm}
    \centering
    \includegraphics[width=\linewidth]{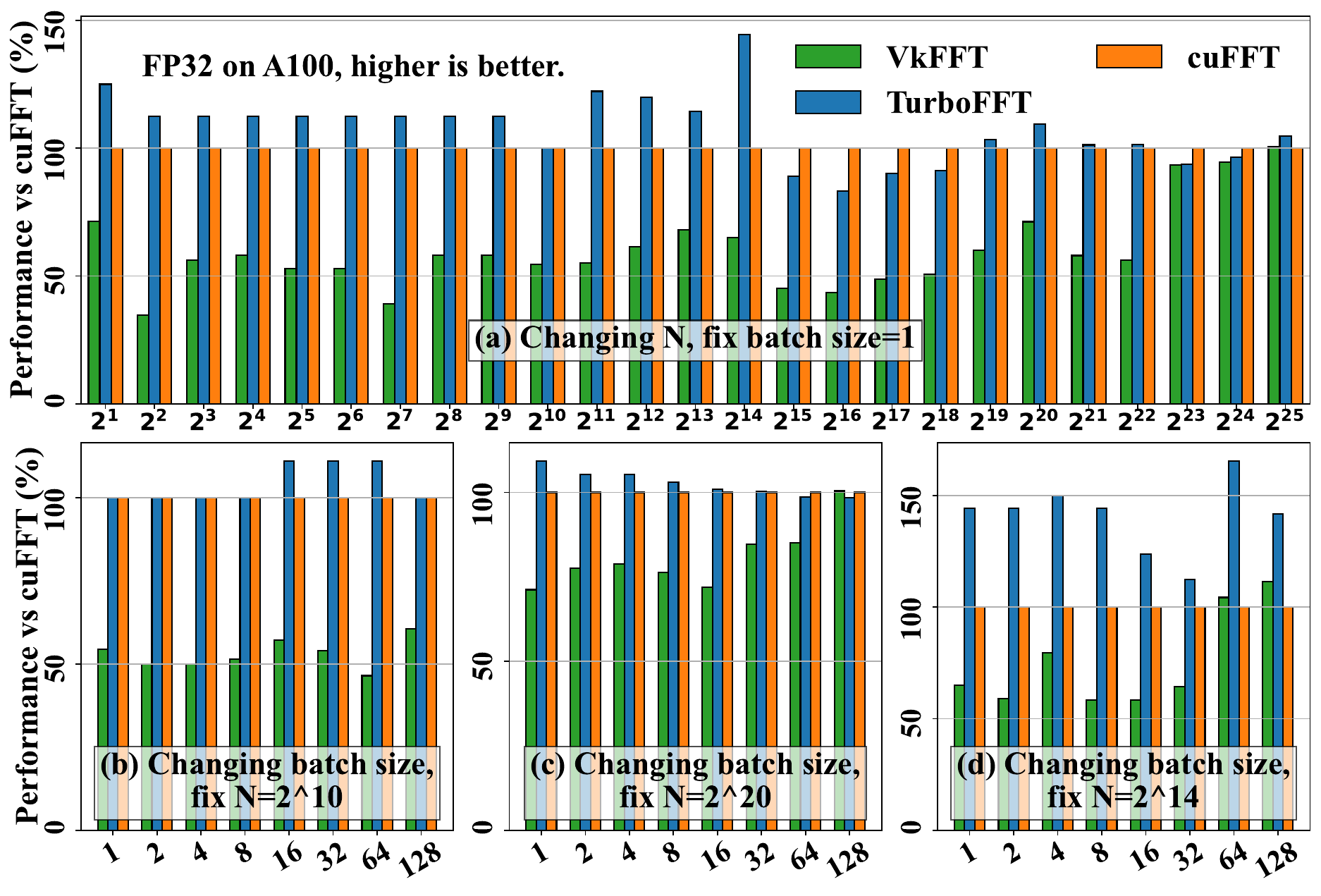}
    \caption{A100 FP32.}
    \label{fig:3xbarchart_overhead_A100_FP32}
    \vspace{-1mm}
\end{figure}
\textit{Global Memory} Although for smaller FFT sizes, each thread block only needs to launch a small number of threads, when the batch size increases, each thread block should also increase the number of threads launched correspondingly to enhance throughput. For instance, as shown in the diagram, when $\log N$ ranges from 0 to 5, the throughput of the FFT kernel rapidly rises to over 80\% efficiency with the increase in batch size. Global memory demands high cache efficiency. We found that inefficient access methods on the A100 can lead to up to a 100\% loss in L1 cache hit rate, resulting in nearly 10,000 cycles of stall time for each thread block launch. This is especially noticeable in larger FFT computations, such as when $N=2^{24}$. Due to the need for three launches, the kernel in the final launch requires an additional transposition operation, changing the storage in memory from the original $(N_1, N_2, N_3)$ to $(N_3, N_2, N_1)$. Although allocating thread blocks along the $(N_1, N_2)$ direction maximizes data locality, it leads to writing back along the direction with the largest stride $(N2, N1)$, resulting in a significantly high rate of L1-cache misses and a 30\% overhead.

\begin{figure}[ht]
    \vspace{-2mm}
    \centering
    \includegraphics[width=\linewidth]{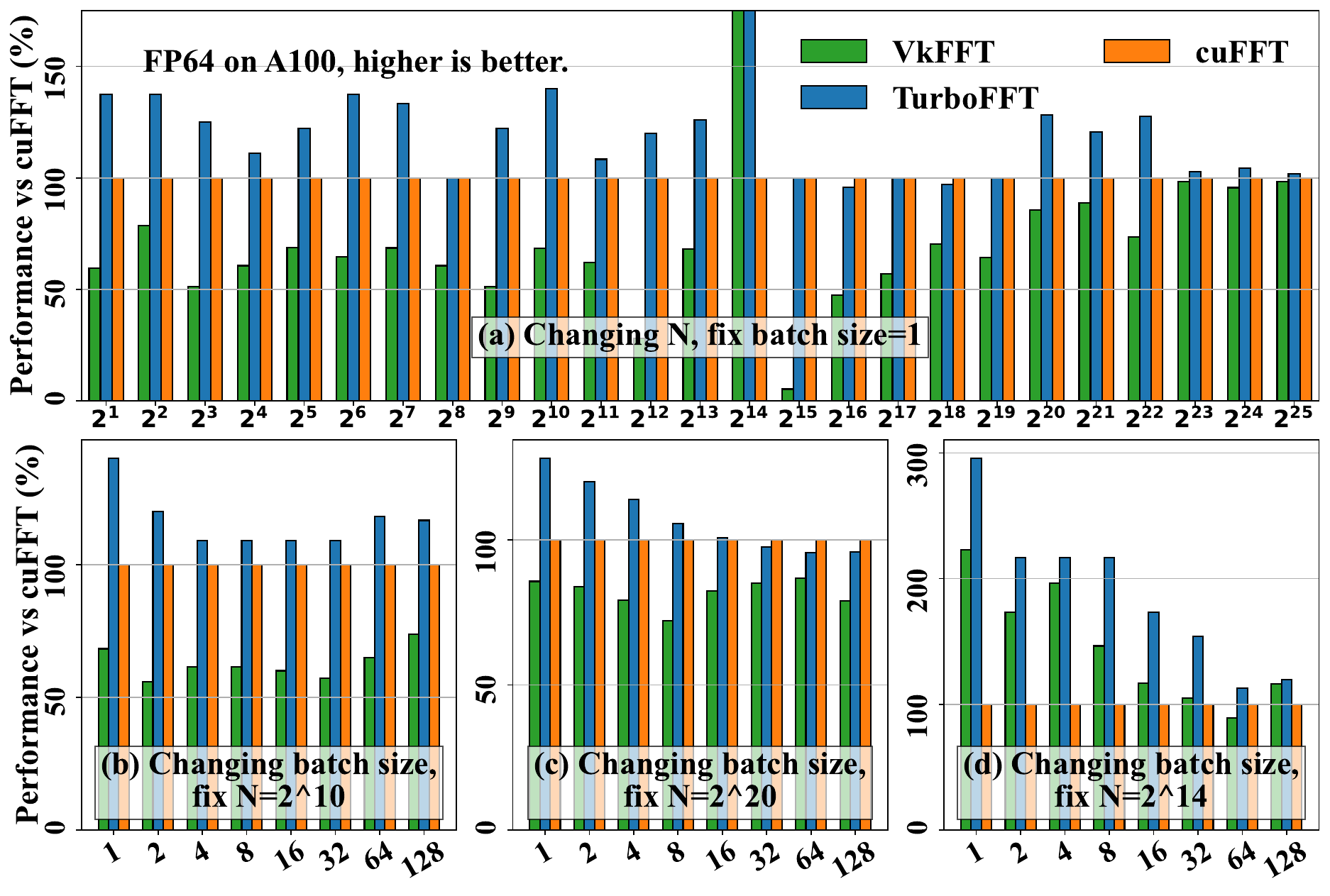}
    \caption{A100 FP64.}
    \label{fig:3xbarchart_overhead_A100_FP64}
    \vspace{-2mm}
\end{figure}
\subsubsection{Padding-Free Design.}
The performance gain of TurboFFT against VkFFT in Figure \ref{fig:3xbarchart_overhead_A100_FP32} (c) and Figure \ref{fig:3xbarchart_overhead_A100_FP64} (c) benefits from the padding-free design. Our padding-free design avoids the bank conflict. The issue of shared memory bank conflict primarily originates from the first twiddling process, where each thread needs to access a continuous memory region, an operation that easily causes different threads within the same warp to access the same memory bank. VkFFT uses padding, namely skipping one bank per 8 or 16 threads. Although this method can avoid bank conflicts, it wastes a significant amount of shared memory, leading to decrement of threadblocks per SM and performance loss in cases with larger N. Padding can be replaced with swizzling. This is feasible thanks to the unit memory transaction size for a single thread of C2C and Z2Z FFTs being 8 bytes or 16 bytes. We observed that this operation yields a 20\% performance improvement when N is small. For FFT sizes that require more than two kernel launches, it's unnecessary to consider bank conflicts. This is because a single warp's threads can be assigned to different batches. By setting the batch\_id as an offset, threads within the same warp can completely avoid the possibility of bank conflicts. In Figure \ref{fig:3xbarchart_overhead_A100_FP32} (c) and Figure \ref{fig:3xbarchart_overhead_A100_FP64} (c), TurboFFT maintains a comparable or superior performance compared to cuFFT. In contrast, VkFFT shows an overhead of more than $20\%$.
\begin{figure}[h]
    \vspace{-2mm}
    \centering
    \includegraphics[width=\linewidth]{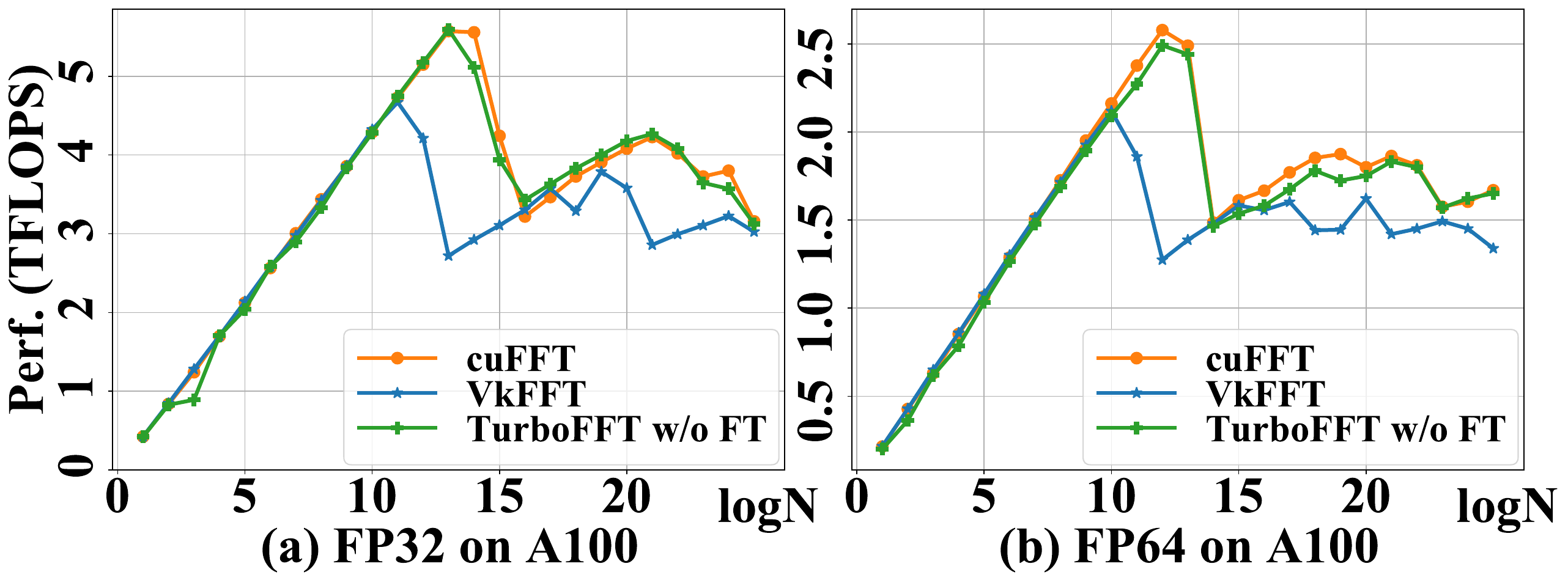}
    \caption{Comparison of batched FFT performance without fault tolerance with TurboFFT, cuFFT and VkFFT on an A100 GPU.}
    \label{fig:a100_bench}
    \vspace{-2mm}
\end{figure}

\subsubsection{Kernel Parameter Selection.}
The superior performance of TurboFFT against cuFFT in Figure \ref{fig:3xbarchart_overhead_A100_FP32} (b)(d) and Figure \ref{fig:3xbarchart_overhead_A100_FP64} (b)(d) benefits from better kernel parameters. The sub-par performance of cuFFT is mainly due to a poor kernel parameter selection. The inefficient kernel parameter cannot fully utilize all 108 streaming multiprocessors in a A100 GPU. In contrast, we manually search for better parameters using template-based code generation. For example, for signal length from $2^7$ to $2^{13}$, each threadblock only performs FFT for one signal, instead of a batch, to maximize streaming processor utilization. Regarding to signal length of $2^{14}$, TurboFFT shows $1.4\times$ to $2\times$ speedup compared to cuFFT for both FP32 and FP64. Our code generation strategy offers flexibility to support a wide range of input shapes and datatypes. 


\begin{figure}[ht]
    \vspace{-2mm}
    \centering
    \includegraphics[width=\linewidth]{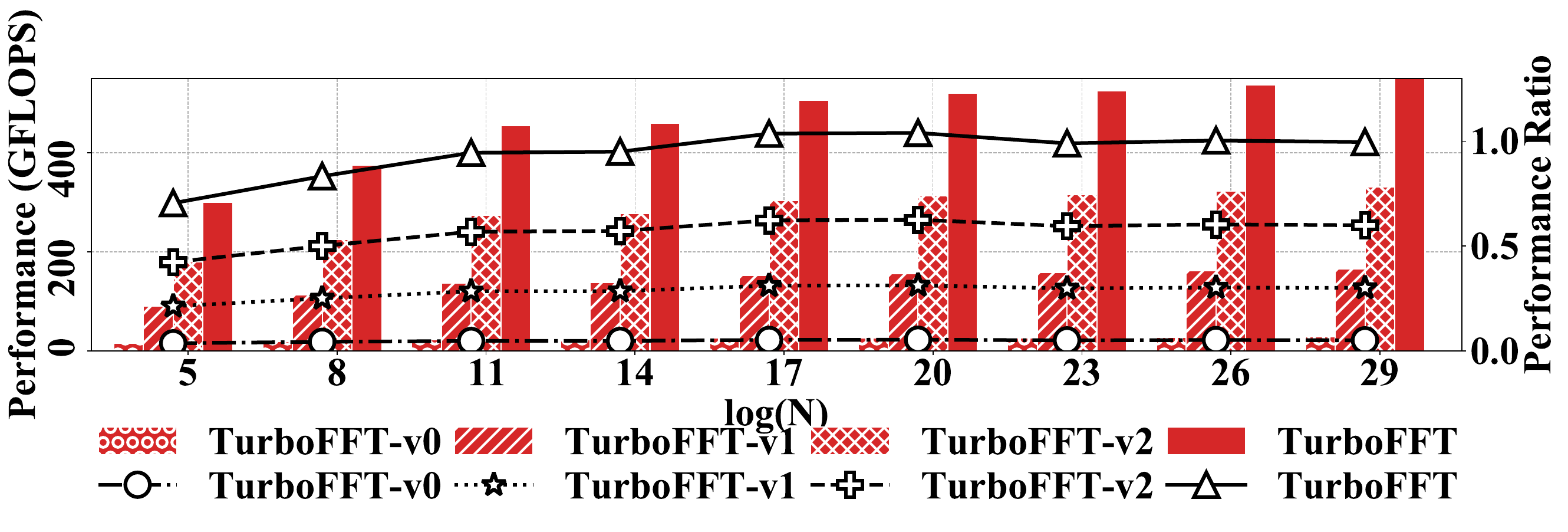}
    \caption{TurboFFT w/o FT stepwise optimizations on T4, FP32}
    \label{fig:stepwise_FP32_t4}
    \vspace{-2mm}
\end{figure}

\subsubsection{Optimiations Impact Overview.}
Figure \ref{fig:stepwise_FP32_t4} illustrates the impact of architecture-aware optimizations, padding-free design, and parameter tuning. The performance is measured with GFLOPS (bar plot, the left y-axis) and the performance ratio with respect to cuFFT (line chart, the right y-axis). For the most basic version, TurboFFT-v0, each thread handles a radix-2 FFT, and a $log_2(N)$ times of kernel launches are required. Without using any optimizations, the TurboFFT-v0 obtains a performance of 49 GFLOPS. Next TurboFFT-v1 employs architecture-aware optimizations and improves the performance from 49 GFLOPS to 110 GFLOPS. Then, TurboFFT-v2 adopts the padding-free design and achieves a performance of 334 GFLOPS. Finally, tuning the kernel parameter accelerates the FFT 565 GFLOPS. By employing the above strategies, we finally obtained an FFT baseline comparable or faster than cuFFT. 
\subsection{TurboFFT with Fault Tolerance}
Figure \ref{fig:twoside_abft_stepwise_optimization_A100_FP32} $(a)\rightarrow(b)$  demonstrates our fault tolerance solution achieves an average overhead of 10\% to 15\% compared to the significant overhead of 30\% to 300\% in offline FT-FFT. Figure \ref{fig:twoside_abft_stepwise_optimization_A100_FP32} $(a)\rightarrow(c)\rightarrow(d)\rightarrow(e)\rightarrow(f)\rightarrow(b)$ also presents the stepwise optimization of our fused fault tolerance in TurboFFT.  The overhead value is documented at each square as well. Figure \ref{fig:twoside_abft_stepwise_optimization_A100_FP64} demonstrates the comparison chain for FP64 on A100. For better understandability, Table \ref{tab:fft_overhead_comparison} lists the method and the comparison base in Figure \ref{fig:twoside_abft_stepwise_optimization_A100_FP32},\ref{fig:twoside_abft_stepwise_optimization_A100_FP64}.
\begin{table}[ht]
\centering
\caption{Method and Comparison Base in Figure \ref{fig:twoside_abft_stepwise_optimization_A100_FP32},\ref{fig:twoside_abft_stepwise_optimization_A100_FP64}.}
\label{tab:fft_overhead_comparison}
\begin{tabular}{|c|l|l|}
\hline
\textbf{Id} & \textbf{Fault Tolerance Method} & \textbf{ Base} \\ \hline
(a) & Offline FT-FFT & cuFFT \\ \hline
(b) & TurboFFT: Optimal among (c)|(d)|(e)|(f) & cuFFT \\ \hline
(c) & TurboFFT: Thread-level & (a) \\ \hline
(d) & TurboFFT: Threadblock-level (TB-1)& (c) \\ \hline
(e) & TurboFFT: Batch-2 dispatch order (TB-2)& (c)|(d)\\ \hline
(f) & TurboFFT: Batch-4 dispatch order (TB-4)& (c)|(d)|(e) \\ \hline
\end{tabular}
\vspace{-5mm}
\end{table}
As shown in Figure \ref{fig:twoside_abft_stepwise_optimization_A100_FP32} $(a) \rightarrow (b)$, the red area decreases substantially, demonstrating the efficiency of our ABFT design. The red area means the fault tolerance cause high overhead compared to cuFFT. TurboFFT with fault tolerance achieves negligible overhead with the help of kernel fusion. Next, we detailed how the negligible overhead is achieved.

\subsubsection{Thread-level ABFT} Thread-level ABFT performs ABFT for thread-level FFTs. The thread-level FFT handles signals with length $2, 4, 8, 16, 32$. Thread-level ABFT only increases computation but introduces zero memory overhead. Figure \ref{fig:twoside_abft_stepwise_optimization_A100_FP32} (c) highlights most data points with green because our thread-level ABFT outperforms the offline FT-FFT with kernel fusion. However, the thread-level ABFT results in redundant computation in all threads. Because FFT requires 2 to 4 threadblock-level synchronizations in each kernel launch, the redundancy in computation will introduce substantial overhead in some cases. Hence, we propose threadblock-level ABFT to decrease the computation overhead. 

\subsubsection{Threadblock-level ABFT} 
\begin{figure}[th]
        \centering
    \includegraphics[width=\linewidth]{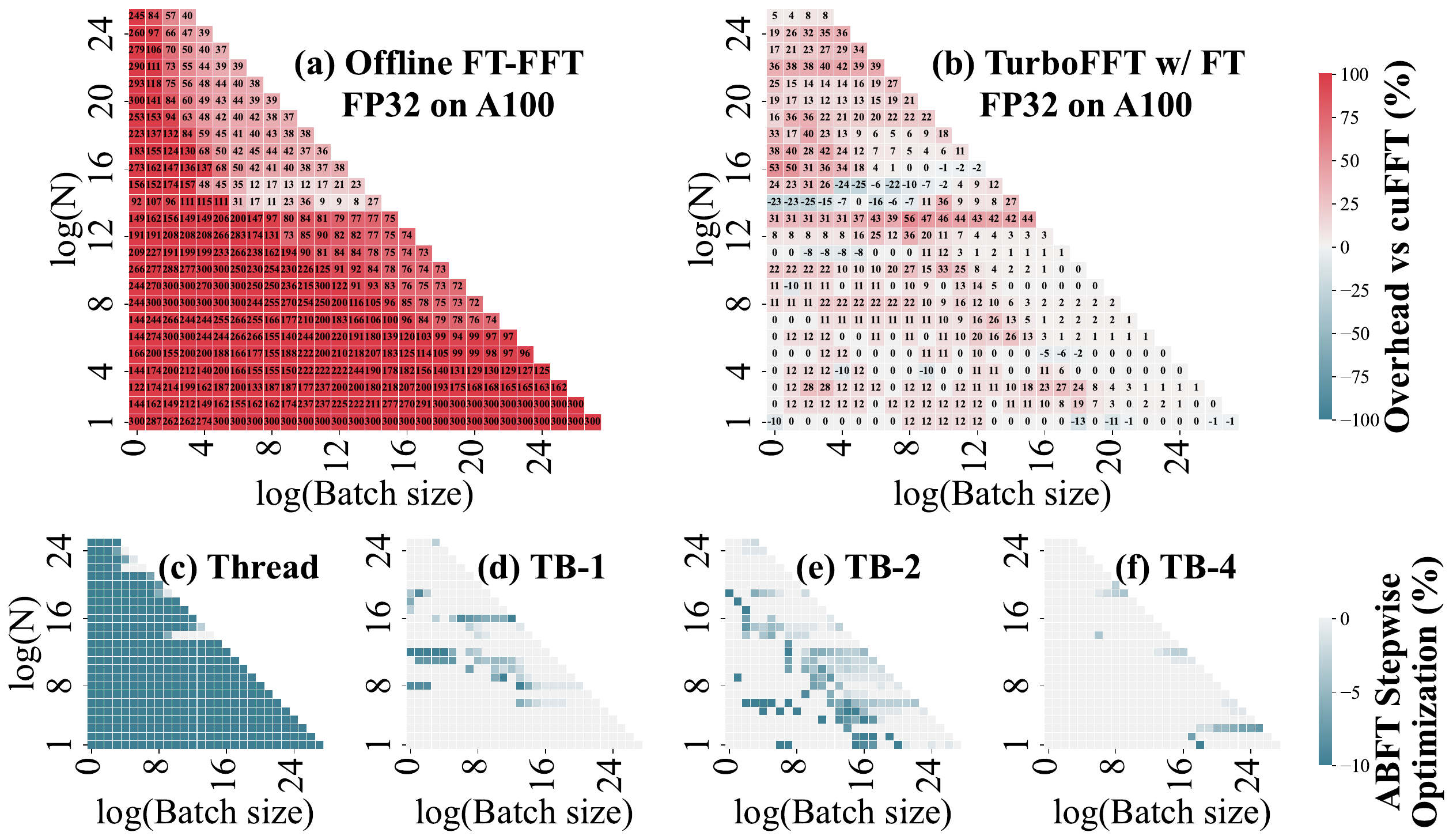} 
        \caption{ABFT stepwise optimization for FP32 FFT on A100.}
        \label{fig:twoside_abft_stepwise_optimization_A100_FP32}
        \vspace{-2mm}
\end{figure}

Threadblock-level ABFT protects the FFT computation at threadblock-level. Those FFTs typically have lengths from $128$ to $2048$. The input and output are assigned to different threads. Hence synchronous reductions are required to compute the checksum. The advantage is the redundant computation in Thread-level ABFT can be removed. Although FFT is typically a memory-bound application, Threadblock-level ABFT still demonstrates performance gain at signal length $2^{12}$ and $2^{16}$, as shown in Figure \ref{fig:twoside_abft_stepwise_optimization_A100_FP32} (d). To further hide the memory footprint, we require each threadblock to continuously perform more than one transaction of \textit{global memory read} $\rightarrow$ \textit{threadblock-level FFT} $\rightarrow$ \textit{global memory write}.

\subsubsection{Multi-Transaction Threadblock-level ABFT} As discussed before, Threadblock-level ABFT results in additional memory overhead. Typically, each threadblock dies once it finishes a threadblock-level FFT transaction, namely \textit{global memory read} $\rightarrow$ \textit{threadblock-level FFT} $\rightarrow$ \textit{global memory write}. This threadblock dispatching pattern results in the checksum that can only be applied to one transaction with a limited parallelism. We can further increase the parallelism of ABFT by requiring one threadblock to perform more than one transaction. Multi-transactions ABFT requires the same number of threadblock-level reductions compared to the original 1-transaction threadblock-level ABFT. Hence the reduction overhead is averaged by the number of transactions. Furthermore, no inter-transaction communication is required! It is because each thread exactly maps to the same ABFT encoding workload, so no additional synchronizations or memory operations are required. The additional computation is only a register-level accumulation of input signals and output signals, which is marginal compared to the FFT original computation. As shown in Figure \ref{fig:twoside_abft_stepwise_optimization_A100_FP32} (e)(f), a 2-transaction threadblock-level ABFT brings a significant improvement compared to 1-transaction threadblock-level ABFT. 4-transaction threadblock-level ABFT is able to further improve the optimal combination of thread-level, threadblock-level, and 2-transaction threadblock-level ABFT. From our experiments, \{8, 16, 32\}-transaction threadblock-level ABFTs bring marginal improvements because the multi-transaction hurts the original FFT global memory access pattern and results in high L1-cache miss rate.  
\begin{figure}[ht]
    \vspace{-2mm}
        \centering
    \includegraphics[width=\linewidth]{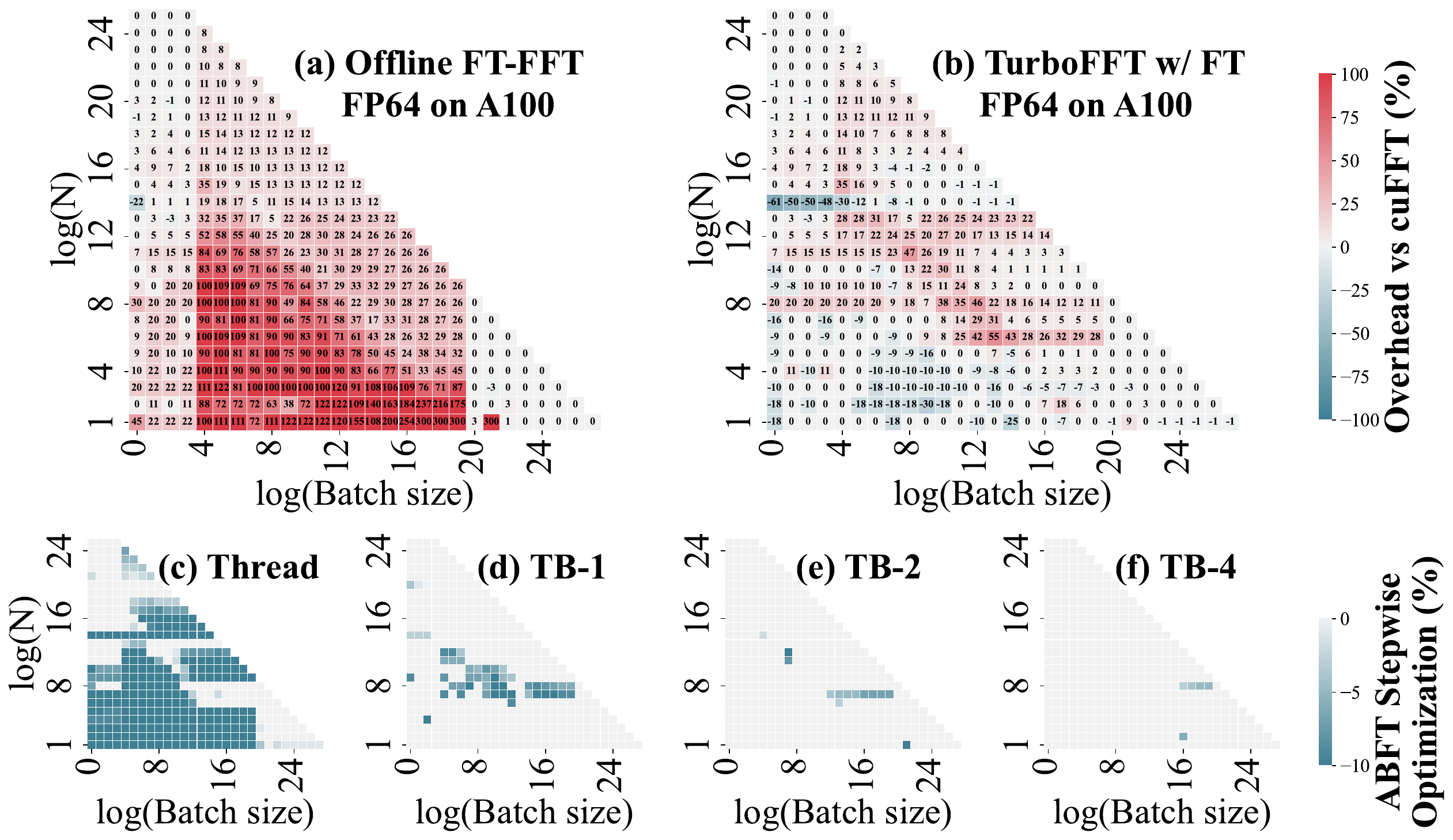} 
        \caption{ABFT stepwise optimization for FP64 FFT on A100.}
        \label{fig:twoside_abft_stepwise_optimization_A100_FP64}
        \vspace{-2mm}
\end{figure}
Figure \ref{fig:twoside_abft_stepwise_optimization_A100_FP64} replicates the experiment for double precision values. Similarly, we observe a sequential decrease in the area covered by red from left to right. Figure \ref{fig:a100_ft} provides a comparison of the performance of TurboFFT in single precision, both without fault tolerance and with fault tolerance, and includes the performance metrics for cuFFT and VkFFT.
\begin{figure}[ht]
\vspace{-2mm}
    \centering
    \includegraphics[width=\linewidth]{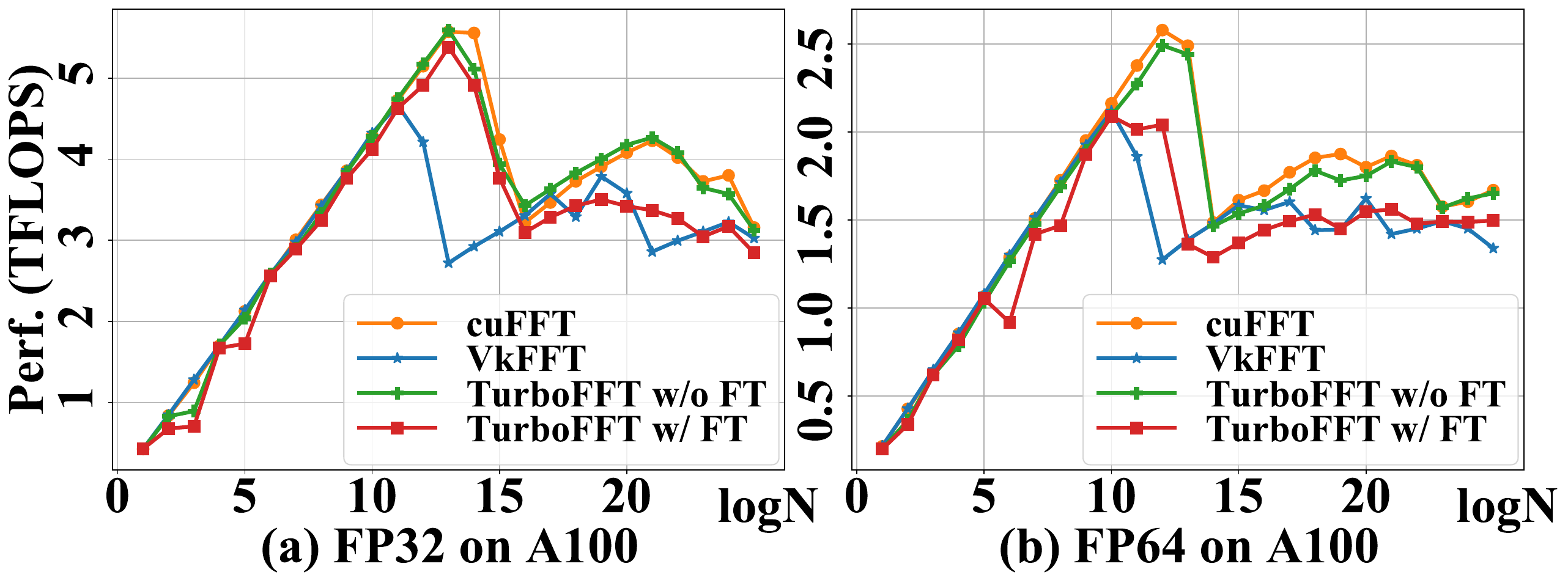}
    \caption{Comparison of TurboFFT performance with and without fault tolerance on an A100 GPU. cuFFT and VkFFT are included. The number of elements is maintained as $2^{28}$.}
\label{fig:a100_ft}
    \vspace{-2mm}
\end{figure}

\subsection{Benchmarking TurboFFT under Error Injection}
\subsubsection{Error Injection Analysis} 
\begin{figure}[ht]
\vspace{-2mm}
    \centering
\includegraphics[scale=0.2]{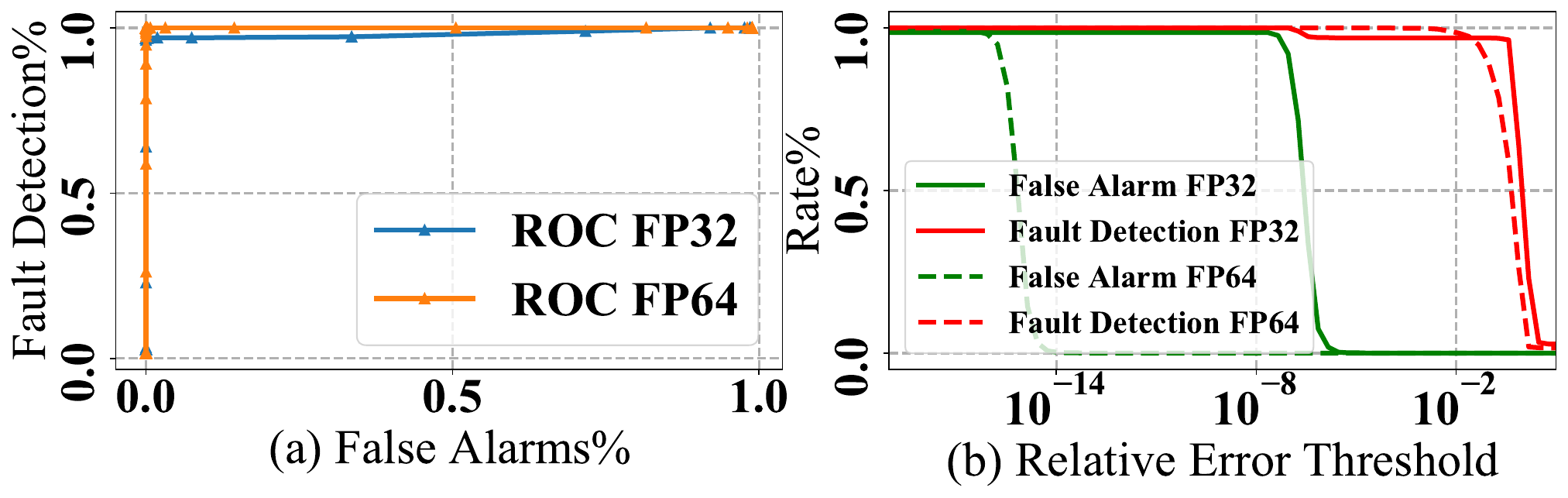} 
    \caption{Error Analysis}
    \label{fig:error_analysis}
    \vspace{-2mm}
\end{figure}

Figure \ref{fig:error_analysis} illustrates that, by selecting an appropriate fault detection threshold $\delta$, the proposed detection scheme is capable of identifying injected faults with a high degree of reliability and a negligible false alarm rate. 2000 random test signals are generated with normal distribution. Faults are injected in half of these runs (1000 of 2000) by first choosing a signal to affect, and then flipping exactly one bit of its 32-bit representation for float-precision and 64-bit representation for double-precision. A checksum test with threshold $\delta$ is used to attempt to identify the affected computations. Characteristics of the proposed fault detection scheme are demonstrated using the standard receiver operating characteristic (ROC) curve in Figure \ref{fig:error_analysis} (a). For a given fault threshold $\delta$, a proportion of False Alarms (numerical errors greater than $\delta$ tagged as data faults) and Detections (injected data faults correctly identified) will be observed. The ROC curve parametrically maps these two proportions as the tolerance level adjusts. Figure \ref{fig:error_analysis} (b) presents the detection rate and false alarm rate versus the fault detection threshold.

\subsubsection{Error Injection Evaluation} 
\begin{figure}[ht]
    \centering
    \includegraphics[width=\linewidth]{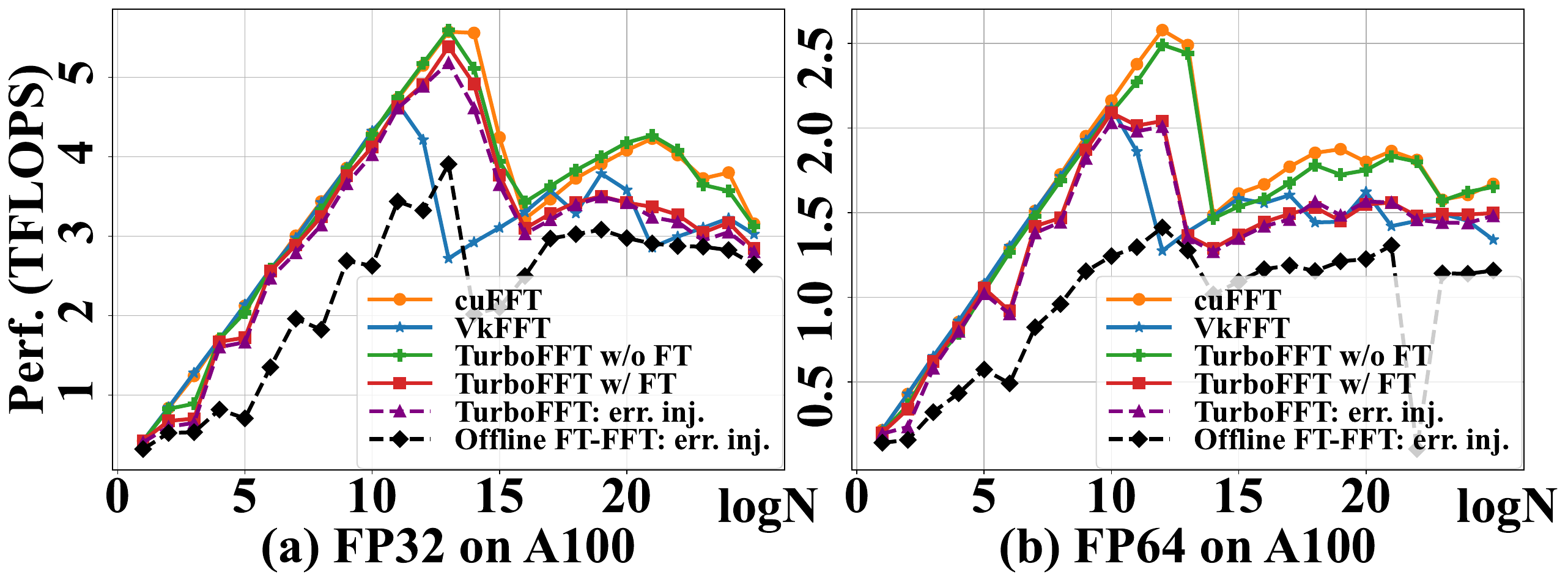}
    \caption{Error injection experiments on A100 GPU}
    \label{fig:err_inj_A100_FP32}
\end{figure}

Figures \ref{fig:err_inj_A100_FP32} extends the analysis to include the performance of TurboFFT under error injection scenarios, and introduces the Offline method for comparison. The figure reveals that TurboFFT when subjected to error injection, incurs a negligible overhead of 3\% for FP32 and 2\% for FP64 compared to scenarios without error injection. Using cuFFT as a baseline, the overhead for TurboFFT with error injection stands at 13\%, whereas the Offline method exhibits a significantly higher overhead of 35\% relative to cuFFT. 

\subsection{Performance Evaluation on T4}
\begin{figure}[ht]
    \centering
\includegraphics[width=1\linewidth]{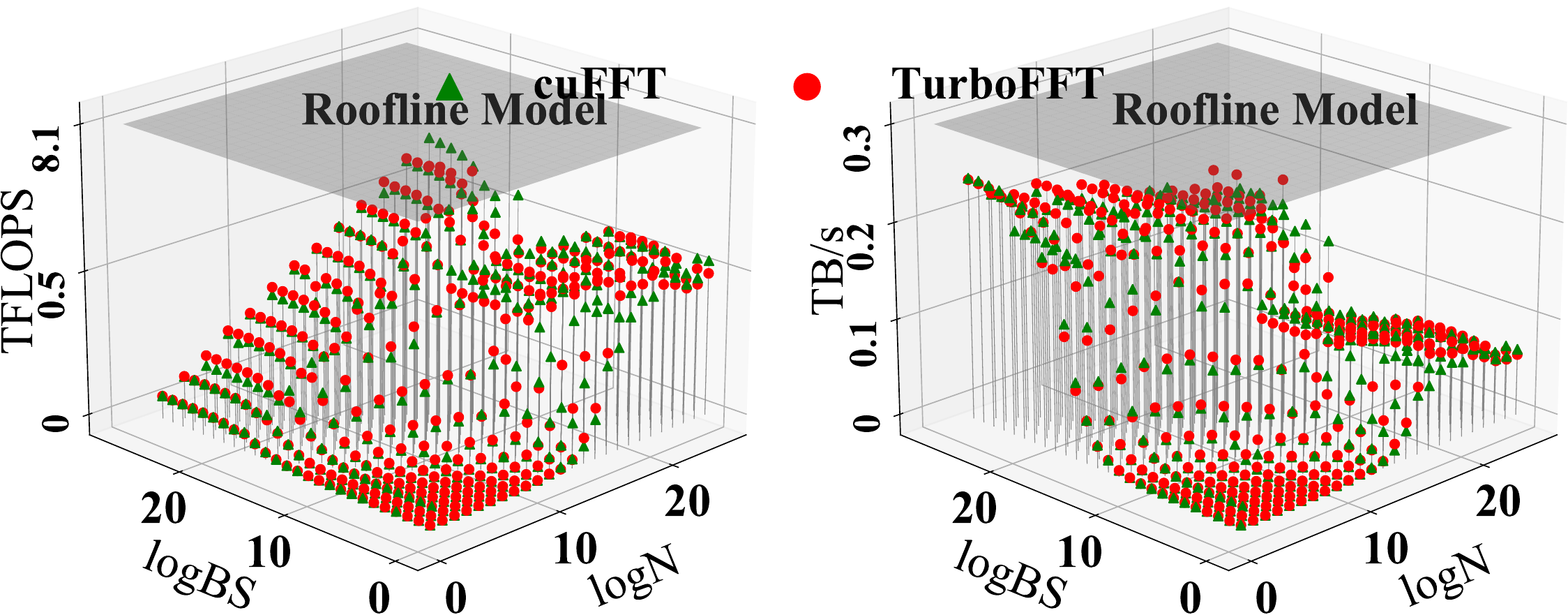} 
    \caption{TurboFFT w/o FT for FP32 on T4.}
    \label{fig:twoside_abft_stepwise_optimization_T4_FP32}
    \vspace{-2mm}
\end{figure}

\begin{figure}[ht]
    \vspace{-2mm}
    \centering
    \includegraphics[width=0.9\linewidth]{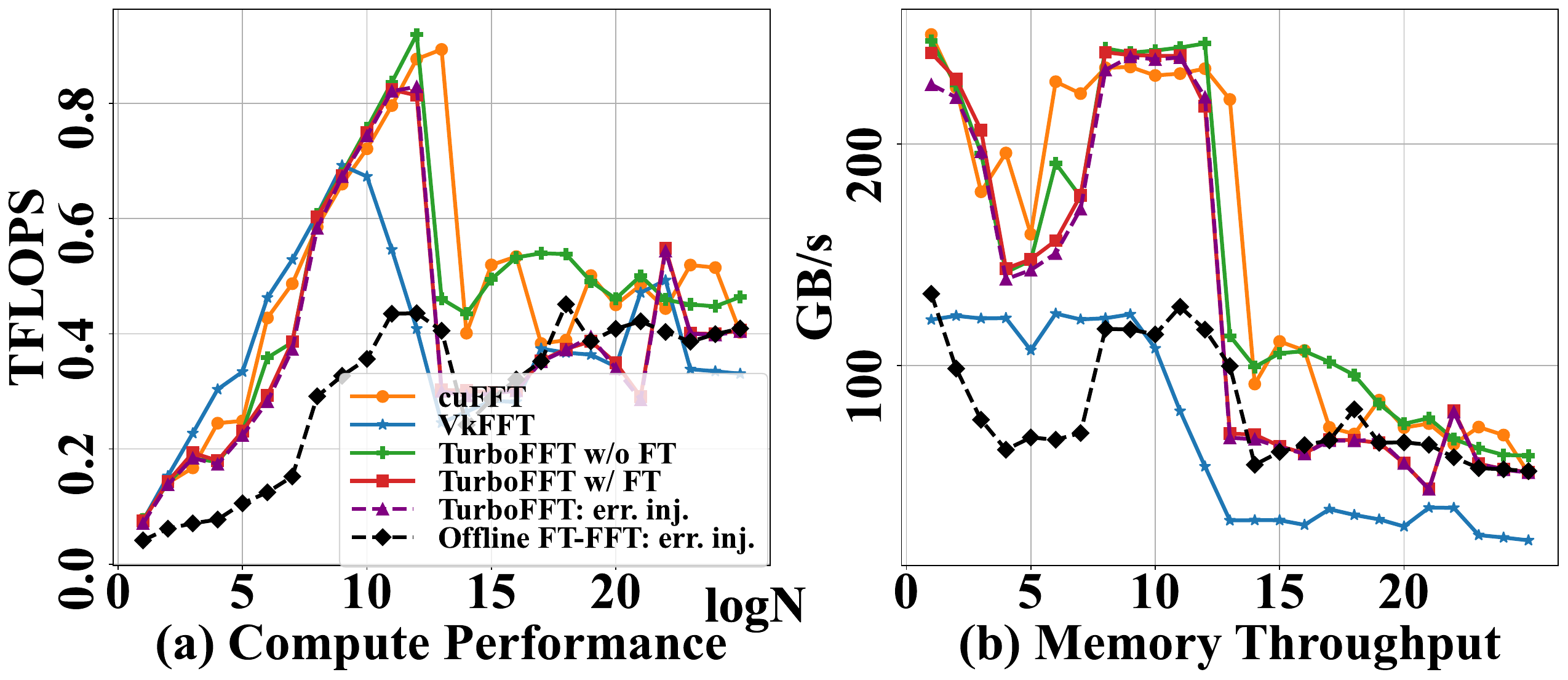}
    \caption{Error injection experiments on T4 GPU}
    \label{fig:err_inj_T4_FP32}
    \vspace{-2mm}
\end{figure}

TurboFFT shows competitive performance for both with or without fault tolerance, as shown in Figure \ref{fig:twoside_abft_stepwise_optimization_T4_FP32} and \ref{fig:err_inj_T4_FP32}. Figure \ref{fig:err_inj_T4_FP32} details the comparison of TurboFFT under error injection on T4, and Offline FT-FFT is included. TurboFFT under error injection incurs a negligible overhead of 3\% for FP32 compared to TurboFFT without error injection. 

\section{Conclusion}
\label{sec:conclusion}
 In this paper, we introduce \textit{TurboFFT}, an FFT prototype codesigned for high-performance and fault tolerance, which not only provides architecture-aware, padding-free and template-based design on FFT side but also presents fused, two-side, multi-transactional ABFT to minimize the fault tolerance overhead. Experimental results on an NVIDIA A100 server GPU and Tesla Turing T4 GPUs show that TurboFFT holds a competitive performance compared to the state-of-the-art closed-sourced library, cuFFT. The fault tolerance scheme in TurboFFT maintains a low overhead (7\% to 15\%), even under hundreds of error injections per minute for both single and double precision. In the future, we would like to extend the fault-tolerant FFT into the scientific compression \cite{liu2024high,jian2024cliz,liu2024cusz,huang2023exploring} and signal processing \cite{johnston2023downlink,johnston2023curriculum,liu2023stationary}.


\bibliographystyle{setting/ACM-Reference-Format}
\bibliography{setting/Reference.bib}


\begin{thebibliography}{54}


\ifx \showCODEN    \undefined \def \showCODEN     #1{\unskip}     \fi
\ifx \showDOI      \undefined \def \showDOI       #1{#1}\fi
\ifx \showISBNx    \undefined \def \showISBNx     #1{\unskip}     \fi
\ifx \showISBNxiii \undefined \def \showISBNxiii  #1{\unskip}     \fi
\ifx \showISSN     \undefined \def \showISSN      #1{\unskip}     \fi
\ifx \showLCCN     \undefined \def \showLCCN      #1{\unskip}     \fi
\ifx \shownote     \undefined \def \shownote      #1{#1}          \fi
\ifx \showarticletitle \undefined \def \showarticletitle #1{#1}   \fi
\ifx \showURL      \undefined \def \showURL       {\relax}        \fi
\providecommand\bibfield[2]{#2}
\providecommand\bibinfo[2]{#2}
\providecommand\natexlab[1]{#1}
\providecommand\showeprint[2][]{arXiv:#2}

\bibitem[Abadi et~al\mbox{.}(2015)]%
        {tensorflow2015-whitepaper}
\bibfield{author}{\bibinfo{person}{Mart\'{\i}n Abadi}, \bibinfo{person}{Ashish Agarwal}, \bibinfo{person}{Paul Barham}, \bibinfo{person}{Eugene Brevdo}, \bibinfo{person}{Zhifeng Chen}, \bibinfo{person}{Craig Citro}, \bibinfo{person}{Greg~S. Corrado}, \bibinfo{person}{Andy Davis}, \bibinfo{person}{Jeffrey Dean}, \bibinfo{person}{Matthieu Devin}, \bibinfo{person}{Sanjay Ghemawat}, \bibinfo{person}{Ian Goodfellow}, \bibinfo{person}{Andrew Harp}, \bibinfo{person}{Geoffrey Irving}, \bibinfo{person}{Michael Isard}, \bibinfo{person}{Yangqing Jia}, \bibinfo{person}{Rafal Jozefowicz}, \bibinfo{person}{Lukasz Kaiser}, \bibinfo{person}{Manjunath Kudlur}, \bibinfo{person}{Josh Levenberg}, \bibinfo{person}{Dandelion Man\'{e}}, \bibinfo{person}{Rajat Monga}, \bibinfo{person}{Sherry Moore}, \bibinfo{person}{Derek Murray}, \bibinfo{person}{Chris Olah}, \bibinfo{person}{Mike Schuster}, \bibinfo{person}{Jonathon Shlens}, \bibinfo{person}{Benoit Steiner}, \bibinfo{person}{Ilya Sutskever}, \bibinfo{person}{Kunal Talwar},
  \bibinfo{person}{Paul Tucker}, \bibinfo{person}{Vincent Vanhoucke}, \bibinfo{person}{Vijay Vasudevan}, \bibinfo{person}{Fernanda Vi\'{e}gas}, \bibinfo{person}{Oriol Vinyals}, \bibinfo{person}{Pete Warden}, \bibinfo{person}{Martin Wattenberg}, \bibinfo{person}{Martin Wicke}, \bibinfo{person}{Yuan Yu}, {and} \bibinfo{person}{Xiaoqiang Zheng}.} \bibinfo{year}{2015}\natexlab{}.
\newblock \bibinfo{title}{{TensorFlow}: Large-Scale Machine Learning on Heterogeneous Systems}.
\newblock
\urldef\tempurl%
\url{https://www.tensorflow.org/}
\showURL{%
\tempurl}
\newblock
\shownote{Software available from tensorflow.org}.


\bibitem[Baumann(2002)]%
        {baumann2002soft}
\bibfield{author}{\bibinfo{person}{Robert Baumann}.} \bibinfo{year}{2002}\natexlab{}.
\newblock \showarticletitle{Soft errors in commercial semiconductor technology: Overview and scaling trends}.
\newblock \bibinfo{journal}{\emph{IEEE 2002 Reliability Physics Tutorial Notes, Reliability Fundamentals}}  \bibinfo{volume}{7} (\bibinfo{year}{2002}).
\newblock


\bibitem[Binder et~al\mbox{.}(1975)]%
        {binder1975satellite}
\bibfield{author}{\bibinfo{person}{Daniel Binder}, \bibinfo{person}{Edward~C Smith}, {and} \bibinfo{person}{AB Holman}.} \bibinfo{year}{1975}\natexlab{}.
\newblock \showarticletitle{Satellite anomalies from galactic cosmic rays}.
\newblock \bibinfo{journal}{\emph{IEEE Transactions on Nuclear Science}} \bibinfo{volume}{22}, \bibinfo{number}{6} (\bibinfo{year}{1975}), \bibinfo{pages}{2675--2680}.
\newblock


\bibitem[Bird et~al\mbox{.}(2017)]%
        {bird2017neutron}
\bibfield{author}{\bibinfo{person}{John~M Bird}, \bibinfo{person}{Michael~K Peters}, \bibinfo{person}{Travis~Z Fullem}, \bibinfo{person}{Michael~J Tostanoski}, \bibinfo{person}{Terrence~F Deaton}, \bibinfo{person}{Kristianto Hartojo}, {and} \bibinfo{person}{Roy~E Strayer}.} \bibinfo{year}{2017}\natexlab{}.
\newblock \showarticletitle{Neutron induced single event upset (seu) testing of commercial memory devices with embedded error correction codes (ecc)}. In \bibinfo{booktitle}{\emph{2017 IEEE Radiation Effects Data Workshop (REDW)}}. IEEE, \bibinfo{pages}{1--8}.
\newblock


\bibitem[Calhoun et~al\mbox{.}(2017)]%
        {calhoun2017towards}
\bibfield{author}{\bibinfo{person}{Jon Calhoun}, \bibinfo{person}{Marc Snir}, \bibinfo{person}{Luke~N Olson}, {and} \bibinfo{person}{William~D Gropp}.} \bibinfo{year}{2017}\natexlab{}.
\newblock \showarticletitle{Towards a more complete understanding of {SDC} propagation}. In \bibinfo{booktitle}{\emph{Proceedings of the 26th International Symposium on High-Performance Parallel and Distributed Computing}}. ACM, \bibinfo{pages}{131--142}.
\newblock


\bibitem[Chen(2008)]%
        {chen2008extending}
\bibfield{author}{\bibinfo{person}{Zizhong Chen}.} \bibinfo{year}{2008}\natexlab{}.
\newblock \showarticletitle{Extending algorithm-based fault tolerance to tolerate fail-stop failures in high performance distributed environments}. In \bibinfo{booktitle}{\emph{2008 IEEE International Symposium on Parallel and Distributed Processing}}. IEEE, \bibinfo{pages}{1--8}.
\newblock


\bibitem[Chen and Dongarra(2008)]%
        {chen2008scalable}
\bibfield{author}{\bibinfo{person}{Zizhong Chen} {and} \bibinfo{person}{Jack Dongarra}.} \bibinfo{year}{2008}\natexlab{}.
\newblock \showarticletitle{A scalable checkpoint encoding algorithm for diskless checkpointing}. In \bibinfo{booktitle}{\emph{2008 11th IEEE High Assurance Systems Engineering Symposium}}. IEEE, \bibinfo{pages}{71--79}.
\newblock


\bibitem[Cher et~al\mbox{.}(2014)]%
        {cher2014understanding}
\bibfield{author}{\bibinfo{person}{Chen-Yong Cher}, \bibinfo{person}{Meeta~S Gupta}, \bibinfo{person}{Pradip Bose}, {and} \bibinfo{person}{K~Paul Muller}.} \bibinfo{year}{2014}\natexlab{}.
\newblock \showarticletitle{Understanding soft error resiliency of blue gene/q compute chip through hardware proton irradiation and software fault injection}. In \bibinfo{booktitle}{\emph{SC'14: Proceedings of the International Conference for High Performance Computing, Networking, Storage and Analysis}}. IEEE, \bibinfo{pages}{587--596}.
\newblock


\bibitem[Ding et~al\mbox{.}(2011)]%
        {ding2011matrix}
\bibfield{author}{\bibinfo{person}{Chong Ding}, \bibinfo{person}{Christer Karlsson}, \bibinfo{person}{Hui Liu}, \bibinfo{person}{Teresa Davies}, {and} \bibinfo{person}{Zizhong Chen}.} \bibinfo{year}{2011}\natexlab{}.
\newblock \showarticletitle{Matrix multiplication on gpus with on-line fault tolerance}. In \bibinfo{booktitle}{\emph{2011 IEEE Ninth International Symposium on Parallel and Distributed Processing with Applications}}. IEEE, \bibinfo{pages}{311--317}.
\newblock


\bibitem[Dixit et~al\mbox{.}(2021)]%
        {dixit2021silent}
\bibfield{author}{\bibinfo{person}{Harish~Dattatraya Dixit}, \bibinfo{person}{Sneha Pendharkar}, \bibinfo{person}{Matt Beadon}, \bibinfo{person}{Chris Mason}, \bibinfo{person}{Tejasvi Chakravarthy}, \bibinfo{person}{Bharath Muthiah}, {and} \bibinfo{person}{Sriram Sankar}.} \bibinfo{year}{2021}\natexlab{}.
\newblock \showarticletitle{Silent data corruptions at scale}.
\newblock \bibinfo{journal}{\emph{arXiv preprint arXiv:2102.11245}} (\bibinfo{year}{2021}).
\newblock


\bibitem[Dongarra et~al\mbox{.}(2011)]%
        {dongarra2011international}
\bibfield{author}{\bibinfo{person}{Jack Dongarra}, \bibinfo{person}{Pete Beckman}, \bibinfo{person}{Terry Moore}, \bibinfo{person}{Patrick Aerts}, \bibinfo{person}{Giovanni Aloisio}, \bibinfo{person}{Jean-Claude Andre}, \bibinfo{person}{David Barkai}, \bibinfo{person}{Jean-Yves Berthou}, \bibinfo{person}{Taisuke Boku}, \bibinfo{person}{Bertrand Braunschweig}, {et~al\mbox{.}}} \bibinfo{year}{2011}\natexlab{}.
\newblock \showarticletitle{The international exascale software project roadmap}.
\newblock \bibinfo{journal}{\emph{International Journal of High Performance Computing Applications}} \bibinfo{volume}{25}, \bibinfo{number}{1} (\bibinfo{year}{2011}), \bibinfo{pages}{3--60}.
\newblock


\bibitem[Fagg and Dongarra(2000)]%
        {fagg2000ft}
\bibfield{author}{\bibinfo{person}{Graham~E Fagg} {and} \bibinfo{person}{Jack~J Dongarra}.} \bibinfo{year}{2000}\natexlab{}.
\newblock \showarticletitle{FT-MPI: Fault tolerant MPI, supporting dynamic applications in a dynamic world}. In \bibinfo{booktitle}{\emph{European parallel virtual machine/message passing interface users’ group meeting}}. Springer, \bibinfo{pages}{346--353}.
\newblock


\bibitem[Fu and Yang(2009)]%
        {fu2009fault}
\bibfield{author}{\bibinfo{person}{Hongyi Fu} {and} \bibinfo{person}{Xuejun Yang}.} \bibinfo{year}{2009}\natexlab{}.
\newblock \showarticletitle{Fault tolerant parallel FFT using parallel failure recovery}. In \bibinfo{booktitle}{\emph{2009 International Conference on Computational Science and Its Applications}}. IEEE, \bibinfo{pages}{257--261}.
\newblock


\bibitem[Geist(2016)]%
        {geist2016supercomputing}
\bibfield{author}{\bibinfo{person}{Al Geist}.} \bibinfo{year}{2016}\natexlab{}.
\newblock \showarticletitle{Supercomputing's monster in the closet}.
\newblock \bibinfo{journal}{\emph{IEEE Spectrum}} \bibinfo{volume}{53}, \bibinfo{number}{3} (\bibinfo{year}{2016}), \bibinfo{pages}{30--35}.
\newblock


\bibitem[Habib et~al\mbox{.}(2016)]%
        {habib2016hacc}
\bibfield{author}{\bibinfo{person}{Salman Habib}, \bibinfo{person}{Adrian Pope}, \bibinfo{person}{Hal Finkel}, \bibinfo{person}{Nicholas Frontiere}, \bibinfo{person}{Katrin Heitmann}, \bibinfo{person}{David Daniel}, \bibinfo{person}{Patricia Fasel}, \bibinfo{person}{Vitali Morozov}, \bibinfo{person}{George Zagaris}, \bibinfo{person}{Tom Peterka}, {et~al\mbox{.}}} \bibinfo{year}{2016}\natexlab{}.
\newblock \showarticletitle{HACC: Simulating sky surveys on state-of-the-art supercomputing architectures}.
\newblock \bibinfo{journal}{\emph{New Astronomy}}  \bibinfo{volume}{42} (\bibinfo{year}{2016}), \bibinfo{pages}{49--65}.
\newblock


\bibitem[Hakkarinen et~al\mbox{.}(2014)]%
        {hakkarinen2014fail}
\bibfield{author}{\bibinfo{person}{Doug Hakkarinen}, \bibinfo{person}{Panruo Wu}, {and} \bibinfo{person}{Zizhong Chen}.} \bibinfo{year}{2014}\natexlab{}.
\newblock \showarticletitle{Fail-stop failure algorithm-based fault tolerance for cholesky decomposition}.
\newblock \bibinfo{journal}{\emph{IEEE Transactions on Parallel and Distributed Systems}} \bibinfo{volume}{26}, \bibinfo{number}{5} (\bibinfo{year}{2014}), \bibinfo{pages}{1323--1335}.
\newblock


\bibitem[Hochschild et~al\mbox{.}(2021)]%
        {hochschild2021cores}
\bibfield{author}{\bibinfo{person}{Peter~H Hochschild}, \bibinfo{person}{Paul Turner}, \bibinfo{person}{Jeffrey~C Mogul}, \bibinfo{person}{Rama Govindaraju}, \bibinfo{person}{Parthasarathy Ranganathan}, \bibinfo{person}{David~E Culler}, {and} \bibinfo{person}{Amin Vahdat}.} \bibinfo{year}{2021}\natexlab{}.
\newblock \showarticletitle{Cores that don't count}. In \bibinfo{booktitle}{\emph{Proceedings of the Workshop on Hot Topics in Operating Systems}}. \bibinfo{pages}{9--16}.
\newblock


\bibitem[Huang et~al\mbox{.}(2023)]%
        {huang2023exploring}
\bibfield{author}{\bibinfo{person}{Jiajun Huang}, \bibinfo{person}{Jinyang Liu}, \bibinfo{person}{Sheng Di}, \bibinfo{person}{Yujia Zhai}, \bibinfo{person}{Zizhe Jian}, \bibinfo{person}{Shixun Wu}, \bibinfo{person}{Kai Zhao}, \bibinfo{person}{Zizhong Chen}, \bibinfo{person}{Yanfei Guo}, {and} \bibinfo{person}{Franck Cappello}.} \bibinfo{year}{2023}\natexlab{}.
\newblock \showarticletitle{Exploring Wavelet Transform Usages for Error-bounded Scientific Data Compression}. In \bibinfo{booktitle}{\emph{2023 IEEE International Conference on Big Data (BigData)}}. IEEE, \bibinfo{pages}{4233--4239}.
\newblock


\bibitem[Jian et~al\mbox{.}(2024)]%
        {jian2024cliz}
\bibfield{author}{\bibinfo{person}{Zizhe Jian}, \bibinfo{person}{Sheng Di}, \bibinfo{person}{Jinyang Liu}, \bibinfo{person}{Kai Zhao}, \bibinfo{person}{Xin Liang}, \bibinfo{person}{Haiying Xu}, \bibinfo{person}{Robert Underwood}, \bibinfo{person}{Shixun Wu}, \bibinfo{person}{Jiajun Huang}, \bibinfo{person}{Zizhong Chen}, {et~al\mbox{.}}} \bibinfo{year}{2024}\natexlab{}.
\newblock \showarticletitle{CliZ: Optimizing lossy compression for climate datasets with adaptive fine-tuned data prediction}. In \bibinfo{booktitle}{\emph{2024 IEEE International Parallel and Distributed Processing Symposium (IPDPS)}}. IEEE, \bibinfo{pages}{417--429}.
\newblock


\bibitem[Johnston et~al\mbox{.}(2023a)]%
        {johnston2023curriculum}
\bibfield{author}{\bibinfo{person}{Jeremy Johnston}, \bibinfo{person}{Xiao-Yang Liu}, \bibinfo{person}{Shixun Wu}, {and} \bibinfo{person}{Xiaodong Wang}.} \bibinfo{year}{2023}\natexlab{a}.
\newblock \showarticletitle{A Curriculum Learning Approach to Optimization with Application to Downlink Beamforming}.
\newblock \bibinfo{journal}{\emph{IEEE Transactions on Signal Processing}} (\bibinfo{year}{2023}).
\newblock


\bibitem[Johnston et~al\mbox{.}(2023b)]%
        {johnston2023downlink}
\bibfield{author}{\bibinfo{person}{Jeremy Johnston}, \bibinfo{person}{Xiao-Yang Liu}, \bibinfo{person}{Shixun Wu}, {and} \bibinfo{person}{Xiaodong Wang}.} \bibinfo{year}{2023}\natexlab{b}.
\newblock \showarticletitle{Downlink Beamforming Optimization via Deep Learning}. In \bibinfo{booktitle}{\emph{2023 59th Annual Allerton Conference on Communication, Control, and Computing (Allerton)}}. IEEE, \bibinfo{pages}{1--5}.
\newblock


\bibitem[Jou and Abraham(1988)]%
        {jou1988fault}
\bibfield{author}{\bibinfo{person}{J-Y Jou} {and} \bibinfo{person}{Jacob~A. Abraham}.} \bibinfo{year}{1988}\natexlab{}.
\newblock \showarticletitle{Fault-tolerant FFT networks}.
\newblock \bibinfo{journal}{\emph{IEEE Trans. Comput.}} \bibinfo{volume}{37}, \bibinfo{number}{5} (\bibinfo{year}{1988}), \bibinfo{pages}{548--561}.
\newblock


\bibitem[Kim et~al\mbox{.}(2018)]%
        {kim2018qmcpack}
\bibfield{author}{\bibinfo{person}{Jeongnim Kim}, \bibinfo{person}{Andrew~D Baczewski}, \bibinfo{person}{Todd~D Beaudet}, \bibinfo{person}{Anouar Benali}, \bibinfo{person}{M~Chandler Bennett}, \bibinfo{person}{Mark~A Berrill}, \bibinfo{person}{Nick~S Blunt}, \bibinfo{person}{Edgar Josu{\'e}~Landinez Borda}, \bibinfo{person}{Michele Casula}, \bibinfo{person}{David~M Ceperley}, {et~al\mbox{.}}} \bibinfo{year}{2018}\natexlab{}.
\newblock \showarticletitle{QMCPACK: an open source ab initio quantum Monte Carlo package for the electronic structure of atoms, molecules and solids}.
\newblock \bibinfo{journal}{\emph{Journal of Physics: Condensed Matter}} \bibinfo{volume}{30}, \bibinfo{number}{19} (\bibinfo{year}{2018}), \bibinfo{pages}{195901}.
\newblock


\bibitem[Laprie(1985)]%
        {laprie1985dependable}
\bibfield{author}{\bibinfo{person}{Jean-Claude Laprie}.} \bibinfo{year}{1985}\natexlab{}.
\newblock \showarticletitle{Dependable computing and fault-tolerance}.
\newblock \bibinfo{journal}{\emph{Digest of Papers FTCS-15}} (\bibinfo{year}{1985}), \bibinfo{pages}{2--11}.
\newblock


\bibitem[Li et~al\mbox{.}(2017)]%
        {li2017understanding}
\bibfield{author}{\bibinfo{person}{Guanpeng Li}, \bibinfo{person}{Siva Kumar~Sastry Hari}, \bibinfo{person}{Michael Sullivan}, \bibinfo{person}{Timothy Tsai}, \bibinfo{person}{Karthik Pattabiraman}, \bibinfo{person}{Joel Emer}, {and} \bibinfo{person}{Stephen~W Keckler}.} \bibinfo{year}{2017}\natexlab{}.
\newblock \showarticletitle{Understanding error propagation in deep learning neural network (DNN) accelerators and applications}. In \bibinfo{booktitle}{\emph{Proceedings of the International Conference for High Performance Computing, Networking, Storage and Analysis}}. \bibinfo{pages}{1--12}.
\newblock


\bibitem[Liang et~al\mbox{.}(2017)]%
        {liang2017correcting}
\bibfield{author}{\bibinfo{person}{Xin Liang}, \bibinfo{person}{Jieyang Chen}, \bibinfo{person}{Dingwen Tao}, \bibinfo{person}{Sihuan Li}, \bibinfo{person}{Panruo Wu}, \bibinfo{person}{Hongbo Li}, \bibinfo{person}{Kaiming Ouyang}, \bibinfo{person}{Yuanlai Liu}, \bibinfo{person}{Fengguang Song}, {and} \bibinfo{person}{Zizhong Chen}.} \bibinfo{year}{2017}\natexlab{}.
\newblock \showarticletitle{Correcting soft errors online in fast {F}ourier transform}. In \bibinfo{booktitle}{\emph{Proceedings of the International Conference for High Performance Computing, Networking, Storage and Analysis}}. ACM, \bibinfo{pages}{30}.
\newblock


\bibitem[Liu et~al\mbox{.}(2024a)]%
        {liu2024high}
\bibfield{author}{\bibinfo{person}{Jinyang Liu}, \bibinfo{person}{Sheng Di}, \bibinfo{person}{Kai Zhao}, \bibinfo{person}{Xin Liang}, \bibinfo{person}{Sian Jin}, \bibinfo{person}{Zizhe Jian}, \bibinfo{person}{Jiajun Huang}, \bibinfo{person}{Shixun Wu}, \bibinfo{person}{Zizhong Chen}, {and} \bibinfo{person}{Franck Cappello}.} \bibinfo{year}{2024}\natexlab{a}.
\newblock \showarticletitle{High-performance Effective Scientific Error-bounded Lossy Compression with Auto-tuned Multi-component Interpolation}.
\newblock \bibinfo{journal}{\emph{Proceedings of the ACM on Management of Data}} \bibinfo{volume}{2}, \bibinfo{number}{1} (\bibinfo{year}{2024}), \bibinfo{pages}{1--27}.
\newblock


\bibitem[Liu et~al\mbox{.}(2024b)]%
        {liu2024cusz}
\bibfield{author}{\bibinfo{person}{Jinyang Liu}, \bibinfo{person}{Jiannan Tian}, \bibinfo{person}{Shixun Wu}, \bibinfo{person}{Sheng Di}, \bibinfo{person}{Boyuan Zhang}, \bibinfo{person}{Robert Underwood}, \bibinfo{person}{Yafan Huang}, \bibinfo{person}{Jiajun Huang}, \bibinfo{person}{Kai Zhao}, \bibinfo{person}{Guanpeng Li}, {et~al\mbox{.}}} \bibinfo{year}{2024}\natexlab{b}.
\newblock \showarticletitle{CUSZ-i: High-Ratio Scientific Lossy Compression on GPUs with Optimized Multi-Level Interpolation}. In \bibinfo{booktitle}{\emph{2024 SC24: International Conference for High Performance Computing, Networking, Storage and Analysis SC}}. IEEE Computer Society, \bibinfo{pages}{158--172}.
\newblock


\bibitem[Liu et~al\mbox{.}(2023)]%
        {liu2023stationary}
\bibfield{author}{\bibinfo{person}{Xiao-Yang Liu}, \bibinfo{person}{Zechu Li}, \bibinfo{person}{Shixun Wu}, {and} \bibinfo{person}{Xiaodong Wang}.} \bibinfo{year}{2023}\natexlab{}.
\newblock \showarticletitle{Stationary deep reinforcement learning with quantum k-spin hamiltonian regularization}. In \bibinfo{booktitle}{\emph{ICLR 2023 Workshop on Physics for Machine Learning}}.
\newblock


\bibitem[Lucas et~al\mbox{.}(2014)]%
        {lucas2014doe}
\bibfield{author}{\bibinfo{person}{Robert Lucas}, \bibinfo{person}{James Ang}, \bibinfo{person}{Keren Bergman}, \bibinfo{person}{Shekhar Borkar}, \bibinfo{person}{William Carlson}, \bibinfo{person}{Laura Carrington}, \bibinfo{person}{George Chiu}, \bibinfo{person}{Robert Colwell}, \bibinfo{person}{William Dally}, \bibinfo{person}{Jack Dongarra}, {et~al\mbox{.}}} \bibinfo{year}{2014}\natexlab{}.
\newblock \bibinfo{booktitle}{\emph{{DOE} advanced scientific computing advisory subcommittee ({ASCAC}) report: top ten exascale research challenges}}.
\newblock \bibinfo{type}{{T}echnical {R}eport}. \bibinfo{institution}{USDOE Office of Science (SC)(United States)}.
\newblock


\bibitem[Lutz(1993)]%
        {lutz1993analyzing}
\bibfield{author}{\bibinfo{person}{Robyn~R Lutz}.} \bibinfo{year}{1993}\natexlab{}.
\newblock \showarticletitle{Analyzing software requirements errors in safety-critical, embedded systems}. In \bibinfo{booktitle}{\emph{[1993] Proceedings of the IEEE International Symposium on Requirements Engineering}}. IEEE, \bibinfo{pages}{126--133}.
\newblock


\bibitem[May and Woods(1979)]%
        {may1979alpha}
\bibfield{author}{\bibinfo{person}{Timothy~C May} {and} \bibinfo{person}{Murray~H Woods}.} \bibinfo{year}{1979}\natexlab{}.
\newblock \showarticletitle{Alpha-particle-induced soft errors in dynamic memories}.
\newblock \bibinfo{journal}{\emph{IEEE Transactions on Electron Devices}} \bibinfo{volume}{26}, \bibinfo{number}{1} (\bibinfo{year}{1979}), \bibinfo{pages}{2--9}.
\newblock


\bibitem[Mitra et~al\mbox{.}(2014)]%
        {mitra2014resilience}
\bibfield{author}{\bibinfo{person}{Subhasish Mitra}, \bibinfo{person}{Pradip Bose}, \bibinfo{person}{Eric Cheng}, \bibinfo{person}{Chen-Yong Cher}, \bibinfo{person}{Hyungmin Cho}, \bibinfo{person}{Rajiv Joshi}, \bibinfo{person}{Young~Moon Kim}, \bibinfo{person}{Charles~R Lefurgy}, \bibinfo{person}{Yanjing Li}, \bibinfo{person}{Kenneth~P Rodbell}, {et~al\mbox{.}}} \bibinfo{year}{2014}\natexlab{}.
\newblock \showarticletitle{The resilience wall: Cross-layer solution strategies}. In \bibinfo{booktitle}{\emph{Proceedings of Technical Program-2014 International Symposium on VLSI Technology, Systems and Application (VLSI-TSA)}}. IEEE, \bibinfo{pages}{1--11}.
\newblock


\bibitem[Murphy and Lyon(1998)]%
        {murphy1998ngst}
\bibfield{author}{\bibinfo{person}{T Murphy} {and} \bibinfo{person}{R Lyon}.} \bibinfo{year}{1998}\natexlab{}.
\newblock \showarticletitle{NGST autonomous optical control system}.
\newblock \bibinfo{journal}{\emph{Space Telescope Science Inst}} (\bibinfo{year}{1998}).
\newblock


\bibitem[Nicolaidis(1999)]%
        {nicolaidis1999time}
\bibfield{author}{\bibinfo{person}{Michael Nicolaidis}.} \bibinfo{year}{1999}\natexlab{}.
\newblock \showarticletitle{Time redundancy based soft-error tolerance to rescue nanometer technologies}. In \bibinfo{booktitle}{\emph{Proceedings 17th IEEE VLSI Test Symposium (Cat. No. PR00146)}}. IEEE, \bibinfo{pages}{86--94}.
\newblock


\bibitem[Oliveira et~al\mbox{.}(2017)]%
        {oliveira2017experimental}
\bibfield{author}{\bibinfo{person}{Daniel Oliveira}, \bibinfo{person}{La{\'e}rcio Pilla}, \bibinfo{person}{Nathan DeBardeleben}, \bibinfo{person}{Sean Blanchard}, \bibinfo{person}{Heather Quinn}, \bibinfo{person}{Israel Koren}, \bibinfo{person}{Philippe Navaux}, {and} \bibinfo{person}{Paolo Rech}.} \bibinfo{year}{2017}\natexlab{}.
\newblock \showarticletitle{Experimental and analytical study of {X}eon {P}hi reliability}. In \bibinfo{booktitle}{\emph{Proceedings of the International Conference for High Performance Computing, Networking, Storage and Analysis}}. ACM, \bibinfo{pages}{28}.
\newblock


\bibitem[Paszke et~al\mbox{.}(2019)]%
        {NEURIPS2019_9015}
\bibfield{author}{\bibinfo{person}{Adam Paszke}, \bibinfo{person}{Sam Gross}, \bibinfo{person}{Francisco Massa}, \bibinfo{person}{Adam Lerer}, \bibinfo{person}{James Bradbury}, \bibinfo{person}{Gregory Chanan}, \bibinfo{person}{Trevor Killeen}, \bibinfo{person}{Zeming Lin}, \bibinfo{person}{Natalia Gimelshein}, \bibinfo{person}{Luca Antiga}, \bibinfo{person}{Alban Desmaison}, \bibinfo{person}{Andreas Kopf}, \bibinfo{person}{Edward Yang}, \bibinfo{person}{Zachary DeVito}, \bibinfo{person}{Martin Raison}, \bibinfo{person}{Alykhan Tejani}, \bibinfo{person}{Sasank Chilamkurthy}, \bibinfo{person}{Benoit Steiner}, \bibinfo{person}{Lu Fang}, \bibinfo{person}{Junjie Bai}, {and} \bibinfo{person}{Soumith Chintala}.} \bibinfo{year}{2019}\natexlab{}.
\newblock \showarticletitle{PyTorch: An Imperative Style, High-Performance Deep Learning Library}.
\newblock In \bibinfo{booktitle}{\emph{Advances in Neural Information Processing Systems 32}}, \bibfield{editor}{\bibinfo{person}{H.~Wallach}, \bibinfo{person}{H.~Larochelle}, \bibinfo{person}{A.~Beygelzimer}, \bibinfo{person}{F.~d\textquotesingle Alch\'{e}-Buc}, \bibinfo{person}{E.~Fox}, {and} \bibinfo{person}{R.~Garnett}} (Eds.). \bibinfo{publisher}{Curran Associates, Inc.}, \bibinfo{pages}{8024--8035}.
\newblock


\bibitem[Petersen et~al\mbox{.}(2013)]%
        {petersen2013single}
\bibfield{author}{\bibinfo{person}{EL Petersen}, \bibinfo{person}{R Koga}, \bibinfo{person}{MA Shoga}, \bibinfo{person}{JC Pickel}, {and} \bibinfo{person}{WE Price}.} \bibinfo{year}{2013}\natexlab{}.
\newblock \showarticletitle{The single event revolution}.
\newblock \bibinfo{journal}{\emph{IEEE Transactions on Nuclear Science}} \bibinfo{volume}{60}, \bibinfo{number}{3} (\bibinfo{year}{2013}), \bibinfo{pages}{1824--1835}.
\newblock


\bibitem[Phillips et~al\mbox{.}(2005)]%
        {phillips2005scalable}
\bibfield{author}{\bibinfo{person}{James~C Phillips}, \bibinfo{person}{Rosemary Braun}, \bibinfo{person}{Wei Wang}, \bibinfo{person}{James Gumbart}, \bibinfo{person}{Emad Tajkhorshid}, \bibinfo{person}{Elizabeth Villa}, \bibinfo{person}{Christophe Chipot}, \bibinfo{person}{Robert~D Skeel}, \bibinfo{person}{Laxmikant Kale}, {and} \bibinfo{person}{Klaus Schulten}.} \bibinfo{year}{2005}\natexlab{}.
\newblock \showarticletitle{Scalable molecular dynamics with {NAMD}}.
\newblock \bibinfo{journal}{\emph{Journal of computational chemistry}} \bibinfo{volume}{26}, \bibinfo{number}{16} (\bibinfo{year}{2005}), \bibinfo{pages}{1781--1802}.
\newblock


\bibitem[Pilla et~al\mbox{.}(2014)]%
        {pilla2014software}
\bibfield{author}{\bibinfo{person}{Laercio~L Pilla}, \bibinfo{person}{Paolo Rech}, \bibinfo{person}{Francesco Silvestri}, \bibinfo{person}{Christopher Frost}, \bibinfo{person}{Philippe Olivier~Alexandre Navaux}, \bibinfo{person}{M~Sonza Reorda}, {and} \bibinfo{person}{Luigi Carro}.} \bibinfo{year}{2014}\natexlab{}.
\newblock \showarticletitle{Software-based hardening strategies for neutron sensitive FFT algorithms on GPUs}.
\newblock \bibinfo{journal}{\emph{IEEE Transactions on Nuclear Science}} \bibinfo{volume}{61}, \bibinfo{number}{4} (\bibinfo{year}{2014}), \bibinfo{pages}{1874--1880}.
\newblock


\bibitem[Reis et~al\mbox{.}(2005)]%
        {reis2005swift}
\bibfield{author}{\bibinfo{person}{George~A Reis}, \bibinfo{person}{Jonathan Chang}, \bibinfo{person}{Neil Vachharajani}, \bibinfo{person}{Ram Rangan}, {and} \bibinfo{person}{David~I August}.} \bibinfo{year}{2005}\natexlab{}.
\newblock \showarticletitle{SWIFT: Software implemented fault tolerance}. In \bibinfo{booktitle}{\emph{Proceedings of the international symposium on Code generation and optimization}}. IEEE Computer Society, \bibinfo{pages}{243--254}.
\newblock


\bibitem[Snir et~al\mbox{.}(2014)]%
        {snir2014addressing}
\bibfield{author}{\bibinfo{person}{Marc Snir}, \bibinfo{person}{Robert~W Wisniewski}, \bibinfo{person}{Jacob~A Abraham}, \bibinfo{person}{Sarita~V Adve}, \bibinfo{person}{Saurabh Bagchi}, \bibinfo{person}{Pavan Balaji}, \bibinfo{person}{Jim Belak}, \bibinfo{person}{Pradip Bose}, \bibinfo{person}{Franck Cappello}, \bibinfo{person}{Bill Carlson}, {et~al\mbox{.}}} \bibinfo{year}{2014}\natexlab{}.
\newblock \showarticletitle{Addressing failures in exascale computing}.
\newblock \bibinfo{journal}{\emph{The International Journal of High Performance Computing Applications}} \bibinfo{volume}{28}, \bibinfo{number}{2} (\bibinfo{year}{2014}), \bibinfo{pages}{129--173}.
\newblock


\bibitem[Stockman and Mather(1999)]%
        {stockman1999ngst}
\bibfield{author}{\bibinfo{person}{HS Stockman} {and} \bibinfo{person}{John Mather}.} \bibinfo{year}{1999}\natexlab{}.
\newblock \showarticletitle{NGST: Seeing the first stars and galaxies form}. In \bibinfo{booktitle}{\emph{Symposium-International Astronomical Union}}, Vol.~\bibinfo{volume}{186}. Cambridge University Press, \bibinfo{pages}{493--499}.
\newblock


\bibitem[Tao et~al\mbox{.}(2018)]%
        {tao2018improving}
\bibfield{author}{\bibinfo{person}{Dingwen Tao}, \bibinfo{person}{Sheng Di}, \bibinfo{person}{Xin Liang}, \bibinfo{person}{Zizhong Chen}, {and} \bibinfo{person}{Franck Cappello}.} \bibinfo{year}{2018}\natexlab{}.
\newblock \showarticletitle{Improving performance of iterative methods by lossy checkponting}. In \bibinfo{booktitle}{\emph{Proceedings of the 27th international symposium on high-performance parallel and distributed computing}}. \bibinfo{pages}{52--65}.
\newblock


\bibitem[Thompson et~al\mbox{.}(2022)]%
        {thompson2022lammps}
\bibfield{author}{\bibinfo{person}{Aidan~P Thompson}, \bibinfo{person}{H~Metin Aktulga}, \bibinfo{person}{Richard Berger}, \bibinfo{person}{Dan~S Bolintineanu}, \bibinfo{person}{W~Michael Brown}, \bibinfo{person}{Paul~S Crozier}, \bibinfo{person}{Pieter~J In't~Veld}, \bibinfo{person}{Axel Kohlmeyer}, \bibinfo{person}{Stan~G Moore}, \bibinfo{person}{Trung~Dac Nguyen}, {et~al\mbox{.}}} \bibinfo{year}{2022}\natexlab{}.
\newblock \showarticletitle{LAMMPS-a flexible simulation tool for particle-based materials modeling at the atomic, meso, and continuum scales}.
\newblock \bibinfo{journal}{\emph{Computer Physics Communications}}  \bibinfo{volume}{271} (\bibinfo{year}{2022}), \bibinfo{pages}{108171}.
\newblock


\bibitem[Van~Loan(1992)]%
        {van1992computational}
\bibfield{author}{\bibinfo{person}{Charles Van~Loan}.} \bibinfo{year}{1992}\natexlab{}.
\newblock \bibinfo{booktitle}{\emph{Computational frameworks for the fast Fourier transform}}.
\newblock \bibinfo{publisher}{SIAM}.
\newblock


\bibitem[Wang and Jha(1994)]%
        {wang1994algorithm}
\bibfield{author}{\bibinfo{person}{Sying-Jyan Wang} {and} \bibinfo{person}{Niraj~K. Jha}.} \bibinfo{year}{1994}\natexlab{}.
\newblock \showarticletitle{Algorithm-based fault tolerance for FFT networks}.
\newblock \bibinfo{journal}{\emph{IEEE Trans. Comput.}} \bibinfo{volume}{43}, \bibinfo{number}{7} (\bibinfo{year}{1994}), \bibinfo{pages}{849--854}.
\newblock


\bibitem[Wu and Chen(2014)]%
        {wu2014ft}
\bibfield{author}{\bibinfo{person}{Panruo Wu} {and} \bibinfo{person}{Zizhong Chen}.} \bibinfo{year}{2014}\natexlab{}.
\newblock \showarticletitle{FT-ScaLAPACK: Correcting soft errors on-line for {ScaLAPACK Cholesky, QR, and LU} factorization routines}. In \bibinfo{booktitle}{\emph{Proceedings of the 23rd international symposium on High-performance parallel and distributed computing}}. ACM, \bibinfo{pages}{49--60}.
\newblock


\bibitem[Wu et~al\mbox{.}(2024a)]%
        {wu2024ft}
\bibfield{author}{\bibinfo{person}{Shixun Wu}, \bibinfo{person}{Yitong Ding}, \bibinfo{person}{Yujia Zhai}, \bibinfo{person}{Jinyang Liu}, \bibinfo{person}{Jiajun Huang}, \bibinfo{person}{Zizhe Jian}, \bibinfo{person}{Huangliang Dai}, \bibinfo{person}{Sheng Di}, \bibinfo{person}{Bryan~M Wong}, \bibinfo{person}{Zizhong Chen}, {et~al\mbox{.}}} \bibinfo{year}{2024}\natexlab{a}.
\newblock \showarticletitle{FT K-Means: A High-Performance K-Means on GPU with Fault Tolerance}.
\newblock \bibinfo{journal}{\emph{arXiv preprint arXiv:2408.01391}} (\bibinfo{year}{2024}).
\newblock


\bibitem[Wu et~al\mbox{.}(2024b)]%
        {wu2024dgro}
\bibfield{author}{\bibinfo{person}{Shixun Wu}, \bibinfo{person}{Krishnan Raghavan}, \bibinfo{person}{Sheng Di}, \bibinfo{person}{Zizhong Chen}, {and} \bibinfo{person}{Franck Cappello}.} \bibinfo{year}{2024}\natexlab{b}.
\newblock \showarticletitle{DGRO: Diameter-Guided Ring Optimization for Integrated Research Infrastructure Membership}.
\newblock \bibinfo{journal}{\emph{arXiv preprint arXiv:2410.11142}} (\bibinfo{year}{2024}).
\newblock


\bibitem[Wu et~al\mbox{.}(2023a)]%
        {wu2023ft}
\bibfield{author}{\bibinfo{person}{Shixun Wu}, \bibinfo{person}{Yujia Zhai}, \bibinfo{person}{Jiajun Huang}, \bibinfo{person}{Zizhe Jian}, {and} \bibinfo{person}{Zizhong Chen}.} \bibinfo{year}{2023}\natexlab{a}.
\newblock \showarticletitle{Ft-gemm: A fault tolerant high performance gemm implementation on x86 cpus}. In \bibinfo{booktitle}{\emph{Proceedings of the 32nd International Symposium on High-Performance Parallel and Distributed Computing}}. \bibinfo{pages}{323--324}.
\newblock


\bibitem[Wu et~al\mbox{.}(2023b)]%
        {wu2023anatomy}
\bibfield{author}{\bibinfo{person}{Shixun Wu}, \bibinfo{person}{Yujia Zhai}, \bibinfo{person}{Jinyang Liu}, \bibinfo{person}{Jiajun Huang}, \bibinfo{person}{Zizhe Jian}, \bibinfo{person}{Bryan Wong}, {and} \bibinfo{person}{Zizhong Chen}.} \bibinfo{year}{2023}\natexlab{b}.
\newblock \showarticletitle{Anatomy of High-Performance GEMM with Online Fault Tolerance on GPUs}. In \bibinfo{booktitle}{\emph{Proceedings of the 37th International Conference on Supercomputing}}. \bibinfo{pages}{360--372}.
\newblock


\bibitem[Zhai et~al\mbox{.}(2021)]%
        {zhai2021ft}
\bibfield{author}{\bibinfo{person}{Yujia Zhai}, \bibinfo{person}{Elisabeth Giem}, \bibinfo{person}{Quan Fan}, \bibinfo{person}{Kai Zhao}, \bibinfo{person}{Jinyang Liu}, {and} \bibinfo{person}{Zizhong Chen}.} \bibinfo{year}{2021}\natexlab{}.
\newblock \showarticletitle{FT-BLAS: a high performance BLAS implementation with online fault tolerance}. In \bibinfo{booktitle}{\emph{Proceedings of the ACM International Conference on Supercomputing}}. \bibinfo{pages}{127--138}.
\newblock


\bibitem[Zuk et~al\mbox{.}({[n.\,d.]})]%
        {zukswarm}
\bibfield{author}{\bibinfo{person}{Pawel Zuk}, \bibinfo{person}{Hongwei Jin}, \bibinfo{person}{Imtiaz Mahmud}, \bibinfo{person}{Krishnan Raghavan}, \bibinfo{person}{Komal Thareja}, \bibinfo{person}{Shixun Wu}, \bibinfo{person}{Prasanna Balaprakash}, \bibinfo{person}{Franck Cappello}, \bibinfo{person}{Zizhong Chen}, \bibinfo{person}{Ewa Deelman}, {et~al\mbox{.}}} \bibinfo{year}{[n.\,d.]}\natexlab{}.
\newblock \showarticletitle{SWARM: Scientific Workflow Applications on Resilient Metasystem}.
\newblock  (\bibinfo{year}{[n.\,d.]}).
\newblock


\end{thebibliography}

\appendix

\newcommand{\shixun}[1]{\textcolor{blue}{#1}}


\section{Artifact Description}

\subsection{Paper's Main Contributions}



\begin{itemize}

\item[$C_1$] \textbf{TurboFFT without fault tolerance} outperforms the popular open-source library VkFFT, and is comparable to the state-of-the-art closed-source library, cuFFT.

\item[$C_2$] \textbf{TurboFFT with two-side fault tolerance} efficiently fuses the checksum computation into the FFT kernel, minimizing the fault tolerance overhead compared to exisiting offline fault-tolerant FFT (FT-FFT).

\item[$C_3$] \textbf{TurboFFT's online error correction} protects FFT computation on-the-fly, obataining lower error correction overhead compared to the time-redundant recomputation in offline FT-FFT under error injections.

\end{itemize}

\subsection{Computational Artifacts}



Table \ref{tab:contribution} illustrates the relation between TurboFFT's contributions and the experimental figures presented in the paper.
\begin{table}[h]
\caption{Relation between TurboFFT ($A_1$) to $C_{1-3}$.}
\begin{center}
\begin{tabular}{rll}
\toprule
Artifact ID  &  Contributions &  Related \\
             &  Supported     &  Paper Elements \\
\midrule
$A_1$   &  $C_1$ & Figure 1, 10-14, 21 \\
        &  $C_2$      & Figure 16-18, 19-20\\
        &  $C_3$      & Figure 19-20, 22\\
\bottomrule
\end{tabular}
\end{center}
\label{tab:contribution}
\end{table}

\section{Artifact Evaluation}
This document demonstrates how to reproduce the results in the paper: \textbf{TurboFFT: Co-Designed High-Performance and
Fault-Tolerant Fast Fourier Transform on GPUs}.

All supplementary files are available on Zenodo.
The repository \shixun{\texttt{\artifact}} consists of all code, including two \textbf{one-command scripts} \shixun{\texttt{run\_A100.sh}} and \shixun{\texttt{run\_T4.sh}} to reproduce all result figures list in Table \ref{tab:contribution}.

We executed all benchmarks in the paper using the hardware in Table \ref{tab:hardware} and software in Table in \ref{tab:software}.

\begin{table}[h]
\caption{Hardware Environment}
\begin{center}
  \resizebox{\linewidth}{!}{%
\begin{tabular}{rll}
\toprule
System  &  Type &  Description \\
\midrule
System A   & GPU & 1x NVIDIA A100-SXM4-40GB  \\
        &  GPU Power      &  400 W \\
        &  CPU  & AMD EPYC 7713 64-Core Processor \\
        &  Cores per socket  & 64 \\
        &  Threads per cores  & 2 \\
        &  Memory  & 256 GB \\
\midrule
System B   & GPU & 1x NVIDIA Tesla-T4  \\
        &  GPU Power      &  70 W \\
        &  CPU  & Intel(R) Xeon(R) Silver 4216 CPU \\
        &  Cores per socket  & 16 \\
        &  Threads per cores  & 2 \\
        &  Memory  & 192 GB \\
\bottomrule
\end{tabular}
}
\end{center}
\label{tab:hardware}
\end{table}

\begin{table}[h]
\caption{Software Environment}
\begin{center}
\begin{tabular}{rll}
\toprule
System  & Software &  Version \\
\midrule
System A   & \texttt{gcc} & \texttt{12.3.0}  \\
        &  \texttt{cmake}  &  \texttt{3.24.3} \\
        &  \texttt{cudatoolkit}  & \texttt{12.0} \\
        &  \texttt{python}  & \texttt{3.10.14} \\
        &  \texttt{torch}  & \texttt{2.5.1} \\
        &  \texttt{numpy}  & \texttt{2.1.3} \\
        &  \texttt{matplotlib}  & \texttt{3.8.4} \\
        &  \texttt{seaborn}  & \texttt{0.13.2} \\
\midrule
System B   & \texttt{gcc} & \texttt{11.2.0}  \\
        &  \texttt{cmake}  &  \texttt{3.26.4} \\
        &  \texttt{cudatoolkit}  & \texttt{11.6} \\
        &  \texttt{python}  & \texttt{3.9.18} \\
        &  \texttt{torch}  & \texttt{2.5.1} \\
        &  \texttt{numpy}  & \texttt{2.0.2} \\
        &  \texttt{matplotlib}  & \texttt{3.9.2} \\
        &  \texttt{seaborn}  & \texttt{0.13.2} \\
\bottomrule
\end{tabular}
\end{center}
\label{tab:software}
\end{table}

\section{Getting started}\label{sec:getting_started}
This section guides you through the necessary steps to setup your machine. Please follow these steps before starting to reproduce the results.

\subsection{Extract code repositories}
Start by downloading \shixun{\texttt{\artifact}}. Extract the archive into an empty directory
and change into this directory using the following commands. Make sure that the absolute
path to this directory contains no spaces.

\begin{minted}[bgcolor=bgcolor, fontsize=\small]{bash}
# Create a reproduce directory
mkdir reproduce
cd reproduce
cp <path-to>/PPoPP25_Artifact_TurboFFT.zip ./
# Extract the artifact
unzip <path-to>/PPoPP25_Artifact_TurboFFT.zip
cd PPoPP25_Artifact_TurboFFT
\end{minted}

Now, the folder reproduce contains all the necessary code to produce all results shown
in the paper. The code consists of the \texttt{TurboFFT}, and \texttt{Common} repositores.
\texttt{TurboFFT} includes the source code of high-performance FFT library shown in the paper, and the directory
\texttt{Common} contains the cuda helper functions from \texttt{NVIDIA/cuda-samples}.

\subsection{Install host machine compilation prerequisites}
\subsubsection{\texttt{GCC}}
Please install a recent \texttt{gcc} version ($>=$ 11.2.0).
\subsubsection{\texttt{CMake}}
Please install a \texttt{CMake} version ($>=$ 3.24.3).
\subsubsection{\texttt{CUDA Toolkit}}
Please install CUDA Toolkit 12.0 for A100 machine or CUDA
11.6 for the T4 machine.
\begin{minted}[bgcolor=bgcolor, fontsize=\small]{bash}
# Check the version after Installing
gcc --version
cmake --version
nvcc --version
\end{minted}

\subsection{Install host machine codegen \& plot prerequisites}
\subsubsection{\texttt{Python}}
Please install a recent version ($>=$ 3.9).
\subsubsection{\texttt{PyTorch}}
Please install a recent version ($>=$ 2.4).
\subsubsection{\texttt{NumPy}}
Please install a recent version ($>=$ 2.0.2).
\subsubsection{\texttt{Matplotlib}}
Please install a recent version ($>=$ 3.8.4).
\subsubsection{\texttt{Seaborn}}
Please install a recent version ($>=$ 0.13.2).
\begin{minted}[bgcolor=bgcolor, fontsize=\small]{bash}
# Sample instructions to set your python env
python -m venv .venv --prompt turbofft
source .venv/bin/activate
pip install upgrade
pip install torch
pip install numpy
pip install matplotlib
pip install seaborn
\end{minted}

\section{Reproducing Results}

The \textbf{one-command scripts} \shixun{\texttt{run\_A100.sh}} and \shixun{\texttt{run\_T4.sh}} allow you to regenerate \textbf{11 experimental result figures} on NVIDIA A100 GPUs (Figures 1, 10-14, and 16-20) and \textbf{2 experimental result figures} on NVIDIA T4 GPUs (Figures 21-22).

\subsection{How to Run}

\begin{enumerate}
    \item \textbf{Ensure all dependencies are installed} (see Section \ref{sec:getting_started}).

    \item \textbf{Run the script}:
\begin{itemize}
\item On NVIDIA A100 Machine:
\begin{minted}[bgcolor=bgcolor, fontsize=\small]{bash}
./run_A100.sh
\end{minted}  
\item On NVIDIA T4 Machine:
\begin{minted}[bgcolor=bgcolor, fontsize=\small]{bash}
./run_T4.sh
\end{minted}  
\end{itemize}

    \item \textbf{View results}:
    \begin{itemize}
        \item Experimental data will be available in the \shixun{\texttt{TurboFFT/}} \shixun{\texttt{artifact\_data}} directory.
        \item Figures will be saved in the \shixun{\texttt{TurboFFT/}}\shixun{\texttt{artifact\_}} \shixun{\texttt{figures}} directory.
    \end{itemize}
\end{enumerate}

\subsection{Workflow Overview}


The scripts \shixun{\texttt{run\_A100.sh}} and \shixun{\texttt{run\_T4.sh}} execute the following steps:
\begin{enumerate}
    \item \textbf{Environment Setup}: Configures environment variables.
    \item \textbf{Code Generation}: Generates required CUDA kernels.
    \item \textbf{Compilation}: Builds TurboFFT and related binaries.
    \item \textbf{Benchmarking}: Runs benchmarks for TurboFFT.
    \item \textbf{Plotting}: Produces figures matching the paper's results.
\end{enumerate}

\subsection{Runtime Details}

Table \ref{tab:time} shows the estimated execution time of \texttt{run\_A100.sh} and \texttt{run\_T4.sh}. The script \texttt{run\_A100.sh} takes approximately \textbf{2 hours} on a machine with an AMD EPYC 7763 64-Core Processor and a NVIDIA A100 40GB GPU. The script \texttt{run\_T4.sh} takes approximately \textbf{30 minutes} on a machine with an Intel(R) Xeon(R) Silver 4216 CPU and a NVIDIA T4 GPU.

\begin{table}[th]
\caption{Estimated Execution Time}
\begin{center}
  \resizebox{\linewidth}{!}{%
\begin{tabular}{rlllll}
\toprule
System  &  CodeGen &  TurboFFT & Baseline & Plot & Total\\
        &           &           &  cuFFT VkFFT      &  \\
\midrule
\texttt{run\_A100.sh}   & 20 s & 15 min & 90 min  & 3 min & 2 hr \\
                    
\midrule
\texttt{run\_T4.sh}   & 20 s & 10 min & 10 min  & 3 min & 30 min \\

\bottomrule
\end{tabular}
}
\end{center}
\label{tab:time}
\end{table}

\end{document}